\documentclass[aps,pre,review,showpacs,superscriptaddress,12pt]{revtex4-2}
\usepackage{amssymb,amsmath}
\usepackage{amsfonts}
\usepackage{dcolumn}
\usepackage{lipsum}
\usepackage{mathrsfs}
\usepackage{graphicx,subfigure}
\usepackage{color}
\usepackage{comment}
\usepackage{epsfig}
\usepackage{rotating}
\usepackage{tikz}

\newcommand{\iisc}
{\affiliation{Centre for Condensed Matter Theory, Department of Physics, Indian Institute of Science, Bangalore 560012, India}}

\newcommand{\ongil}
{\affiliation{Ongil Private Limited, Chennai 600113, India}}

\newcommand{\A}
{\affiliation{Interdisciplinary Mathematical Sciences, Department of Mathematics, Indian Institute of Science, Bangalore 560012, India}}

\newcommand{\B}
{\affiliation{Bioinformatics Lab, Department of Biochemistry, Indian Institute of Science, Bangalore 560012, India}}

\newcommand{\ibm}
{\affiliation{IBM Research Labs, Bangalore 560045, India}}

\newcommand{\icts}
{\affiliation{International Center for Theoretical Sciences, TIFR, Bangalore 560089, India}}

\begin{document}
\title{Thermodynamics of multi-colored loop models in three dimensions}
\date{\today}

\author{Soumya Kanti Ganguly}%
\email[Email:]{gangulysoumyakanti@gmail.com}
\iisc
\email[Email:]{skg@ongil.ai}
\ongil

\author{Sumanta Mukherjee}%
\email[Email:]{sumantamukherjee@ibm.com}
\A\B\ibm

\author{Chandan Dasgupta}%
\email[Email:]{cdgupta@iisc.ernet.in}
\iisc\icts

\begin{abstract}
We study order-disorder transitions in three-dimensional \textsl{multi-colored} loop models using Monte Carlo simulations. 
We show that the nature of the transition is intimately related to the nature of the loops. The symmetric loops undergo a 
first order phase transition, while the non-symmetric loops show a second-order transition. The critical exponents for the 
non-symmetric loops are calculated. In three dimensions, the regular loop model with no interactions is dual to the XY model. 
We argue that, due to interactions among the colors, the specific heat exponent is found to be different from that of the 
regular loop model. The continuous nature of the transition is altered to a discontinuous one due to the strong inter-color 
interactions.
\end{abstract}

\maketitle

\section{Introduction}

A perfect solid is a regular arrangement of atoms with a given periodicity. The solid-state is characterized by broken
translational and rotational symmetry. However, in reality, perfect solids do not exist but contain irregularities in the
form of vacancies or interstitials. This results in the formation of an extra plane of atoms or a deficiency of the same.
This is a defect which is topological in nature. It means that no amount of smooth deformation of the solid will result 
in the complete removal of the defect. \textsl{Topological defects} that are responsible for the breakdown of the translational 
part of the symmetry are known as dislocations and the ones that break the rotational part are known as disclinations.
A dislocation is a second-rank tensor quantity. One of its indices represents the direction of the \textsl{dislocation line} 
and the other represents the \textsl{Burgers} vector. If the two indices the are same, then it is a \textsl{screw} dislocation; 
otherwise, it is known as an \textsl{edge} dislocation. In general, a dislocated solid will have a combination of these two 
dislocations. For an edge dislocation, the dislocation line along a given direction $i$ and the Burgers vector in the orthogonal 
direction $j$ need not have a symmetric counterpart with the dislocation line along $j$ and the Burgers vector along $i$. Therefore, 
it is reasonable to assume that a dislocation tensor is a second-rank non-symmetric tensor. 

The possibility of dislocations causing melting in solids was first suggested by C. Mott \citep{Mott}. Experiments performed 
by Crawford \citep{Crawford}, Cotterill and Kristensen \citep{Cotterill1} suggest the nucleation of defects in the vicinity of the 
melting transition. This view has been later supported by Molecular dynamics as well as Monte Carlo (MC) simulations \citep{Cotterill2,Janke}.
The first satisfactory theory of the melting transition due to dislocations was put forward by Mizushima \cite{Mizushima} and
Ookawa \cite{Ookawa}, after which many other theories of dislocations and their role in melting were formulated
\citep{Siol,Copeland,Kuhlmann}.

From the statistical physics point of view, dislocations may be treated as extended objects similar to polymers \cite{Edwards} or as 
closed loops \cite{BPS,Vinokur}. As mentioned before, they are second-rank non-symmetric tensors, which obey a continuity condition 
(divergenceless) and have long-range interactions among themselves. These interactions not only depend upon their relative positions but 
also on their orientations \cite{Blin}. Solids in two dimensions transit smoothly (continuous transition) from a less symmetric crystalline 
ordered state to a more symmetric liquid disordered state \cite{KT1,KT2,HalperinNelson1,HalperinNelson2,Young,Strandenburg}. The situation 
is quite the opposite in three dimensions, where the solid-to-liquid (melting) transition or vice-versa is discontinuous in nature. 
MC simulations performed by Janke and Kleinert \cite{JankeKleinert} have confirmed this observation. Their model is essentially 
Gaussian in nature with a combination of elastic and defect degrees of freedom. An alternate viewpoint was suggested by Kleinert 
\cite{Kleinert1,Kleinert2,Kleinert3}, where he looked at the objects that are \textsl{dual} \cite{Kramers,Villain,Savit} to the dislocations 
with short-ranged interactions. These dual objects are divergenceless tensors, which are basically closed loops.

These loops form the basis of our studies reported in this paper and in a companion paper (Ganguly et al. \cite{Ganguly}, hereafter GMDb24). 
The aim of this present study is to further investigate the thermodynamic properties of these loop models using MC simulations.
Since these loops are second-rank tensors ($\bar{\eta}_{ij}$), they can be symmetric ($\bar{\eta}_{ij} = \bar{\eta}_{ji}$) or non-symmetric 
($\bar{\eta}_{ij} \neq \bar{\eta}_{ji}$) in nature. Henceforth, throughout this paper, the symmetric loops will be referred to as SY loops 
and the non-symmetric loops as NS loops. We have considered both the loop models to explore the order-disorder transitions in three-dimensional 
regular lattices. In GMDb24, we focus on analyzing the geometrical properties of these NS loops using certain thermodynamic quantities derived 
herein. In particular, the loop percolation properties of the NS loops have been studied using the finite-size scaling method. The loops obey 
the following divergenceless conditions 
\begin{equation}\label{ZeroDivergence}
\boldsymbol{\Delta}_{i}\bar{\eta}_{ij}(\mathbf{x}) = 0,
\end{equation}
where the index $j = 1,2,3$, will be referred to as the \textsl{color} index (red, blue, green). In both cases, these loops interact via an 
on-site or contact interaction. In other words, these loops can annihilate or reinforce one another only when they share an edge. Our studies show 
that the SY loops undergo a first-order phase transition, whereas a second-order phase transition is seen in the NS loops (see Sec. III). Therefore, 
one can ask whether the nature of the transition is related to the nature of the loops or not. To answer this question, we study the finite temperature 
properties of these loop models using MC simulations. We find that for large negative values of the inter-color interaction parameter ($D$), 
the nature of the transition in NS loops is also altered from a continuous one to a discontinuous one. For different values of $D$ ($0.1,\pm 0.2$, etc.), 
our finite-size scaling studies show that the specific heat exponent ($\alpha$) for the NS loops is different from that of the XY model in three dimensions. 
We argue that this deviation from the second-order behaviour in NS loops is due to the presence of inter-color interactions.

The paper is organized as follows. In Section II, we give a bird's-eye view of Kleinert's theory for dislocations, followed by a detailed description 
of our models. In Section III, we discuss the low and high-temperature properties of both loop models. We then discuss the first-order transition 
and metastability seen in SY loops. Subsequently, we perform the finite-size scaling study of the NS loop system, where we calculate its critical 
exponents. We then look at the relevance of the inter-color coupling in altering the nature of the phase transition in NS loops from a continuous 
transition to a discontinuous one. Finally, in Section IV, we end with some concluding remarks and some plausible future directions. The update rules 
for the SY loops are described in the Appendix.


\section{Models}

In the present section, we briefly discuss Kleinert's theory for dislocations, followed by a detailed description of the 3-color loop models, 
and finally, the methods used to study these loop models. 

\subsection{A brief overview of Kleinert's theory for dislocations}

Motivated by Villain's dual theory for vortices in XY model \cite{Villain}, Kleinert proposed a similar theory for defects
in a lattice \cite{Kleinert1,Kleinert2,Kleinert3,Kleinert4}. Like the non-uniqueness of the phase variables in the XY model,
the displacement of an atom or atoms in a solid is also undetermined up to an integral multiple of the lattice spacing $a$.
If $\theta(\mathbf{x})$ be the value of the phase variable at $\mathbf{x}$, then according to Villain, it remains undetermined
even if $\theta(\mathbf{x}) \rightarrow \theta(\mathbf{x}) + 2n\pi$. Here, $n$ is an integer. Likewise, according to Kleinert,
if $u_{i}$ is the displacement field vector in $i$-th direction, where $i = 1,2,3$, then the displacement of an atom or several
atoms is undetermined for
\begin{equation}\label{Displacement_Fields1A}
\quad u_{i}(\mathbf{x}) \rightarrow u_{i}(\mathbf{x}) + a n_{i}(\mathbf{x}).
\end{equation}
If the lattice derivative of the displacement field vector $u_{j}$ in the $i$-th direction can be written as
\begin{equation}\label{Displacement_Fields1B}
\quad \Delta_{i} u_{j}(\mathbf{x}) = u_{j}(\mathbf{x} + \mathbf{\hat{i}}) - u_{j}(\mathbf{x}),
\end{equation}
then the discrete version of the strain tensor will read as
\begin{equation}\label{Displacement_Fields1C}
u_{ij}(\mathbf{x}) = \Delta_{i} u_{j}(\mathbf{x}) + \Delta_{j} u_{i}(\mathbf{x}).
\end{equation}

The strain tensor will transform like the displacement fields in Eq.\eqref{Displacement_Fields1A} as
\begin{equation}\label{LatticeDerivativeA}
u_{ij}(\mathbf{x}) \rightarrow u_{ij}(\mathbf{x}) + 2a n_{ij}(\mathbf{x}).
\end{equation}
For $i,j = 1,2,3$, the quantities $n_{ij}$ form the singular part of the regular strain fields $u_{ij}$.
Due to the symmetric nature of the strain tensors, the $n_{ij}$'s are also symmetric and integer-valued.
The discrete form of the elastic energy for isotropic solids in terms of these quantities will read as

\begin{equation}\label{ElasticFreeEnergy1A}
E(u_{i}) = \sum_{\mathbf{x}}\bigg[\frac{\mu}{4}\Big(\Delta_{i}u_{j}(\mathbf{x}) + \Delta_{j}u_{i}(\mathbf{x}) \Big)^{2} +
\frac{\lambda}{2} \Big(\Delta_{i} u_{i}(\mathbf{x}) \Big)^{2} \bigg]
\end{equation}

Similar to the XY model, one will have displacement field vectors $u_{i}$ instead of the scalar phase variable. Also, the
scalar vorticity will be replaced by integer-valued tensors $n_{ij}$. Since the partition function for the solids with
$E(u_{i})$ in the exponent is periodic, it may be replaced by a Poisson summation in terms of the integer-valued tensor
variables $\bar{\eta}_{ij}$. For the case $\mu = 2\lambda$, the partition function will have the following formal expression
\begin{eqnarray}\label{PartitionFunction1B}
Z = \sum_{\bar{\eta}_{ij}(\mathbf{x})} \int_{-a/2}^{a/2} \frac{du_{i}(\mathbf{x})}{\sqrt{2\pi/a\lambda\beta}}
\exp\Big[&-&\sum_{\mathbf{x},i,j} \Big(\frac{1}{2a\lambda\beta} \bar{\eta}^{2}_{ij}(\mathbf{x}) \qquad \\ \nonumber
&+& i 2\pi\bar{\eta}_{ij}(\mathbf{x})\boldsymbol{\Delta}_{i}u_{j}(\mathbf{x}) \Big)\Big].
\end{eqnarray}
Where $\beta = 1/k_{B}T$ is the inverse of the temperature ($T$) times the Boltzmann constant. One can now integrate out the 
displacement field variables $u_{i}$ to obtain an expression for the partition function purely in terms of the symmetric 
$\bar{\eta}_{ij}$ variables($\bar{\eta}_{ij} = \bar{\eta}_{ji}$).

\begin{equation}\label{PartitionFunction1C}
Z = \sum_{\bar{\eta}_{ij}(\mathbf{x})} \exp\Big[ -\sum_{\mathbf{x},i,j} \frac{1}{2a\lambda\beta}\bar{\eta}_{ij}^{2}(\mathbf{x}) \Big]
\delta_{\boldsymbol{\Delta}_{i}\bar{\eta}_{ij}(\mathbf{x}),0}
\end{equation}
The Eq.\eqref{PartitionFunction1C} is a restricted sum with the following zero divergence condition
\begin{equation}\label{Divergenceless2A}
\Delta_{i} \bar{\eta}_{ij} = 0,  \quad \forall j = 1,2,3.
\end{equation}
The exponent of the partition function in Eq.\eqref{PartitionFunction1C} is the energy function for the loops. This is a simple quadratic function 
of discrete integer values. Without the continuity condition in Eq.\eqref{Divergenceless2A}, this model is the well-known Discrete Gaussian model 
or the Solid on Substrate (SOS) model \cite{Chui}. 


\subsection{The 3-color loop models}

\begin{figure}
   \centering
   \subfigure[]{\includegraphics[width=0.55\textwidth]{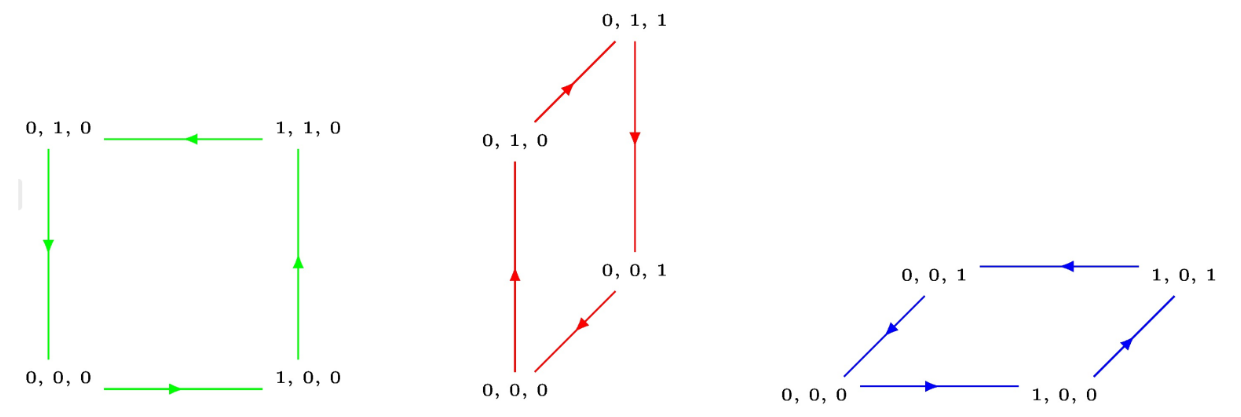}}
   \subfigure[]{\includegraphics[width=0.9\textwidth]{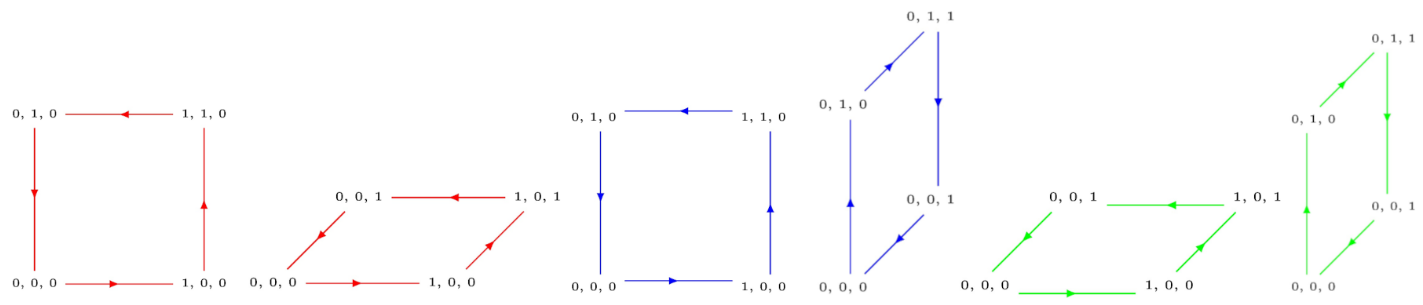}}
   \subfigure[]{\includegraphics[width=0.55\textwidth]{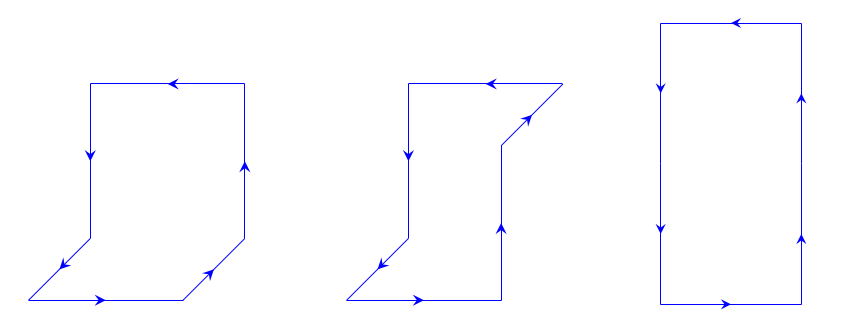}}
   \caption{Elementary excitations in NS-loops: (a) the first excited states with 1 unit of energy ($k_{B}T$ units) and degeneracy 6; 
	(b) the second excited states with 1.5 units of energy and degeneracy 6; (c) shows the third excited state (left and middle) 
	with 2 units of energy and the fourth excited state (right) with 2.5 units of energy, respectively.}
   \label{NSLoopExcitations}
\end{figure}

\begin{figure}
    \centering
    \subfigure[]{\includegraphics[width=0.92\textwidth]{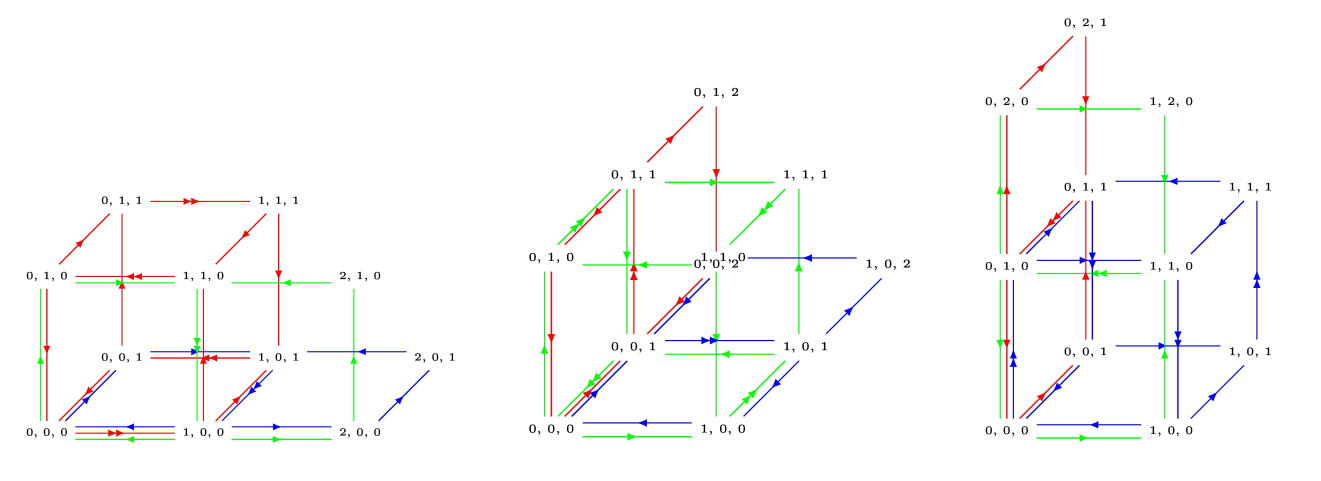}}
    \subfigure[]{\includegraphics[width=0.92\textwidth]{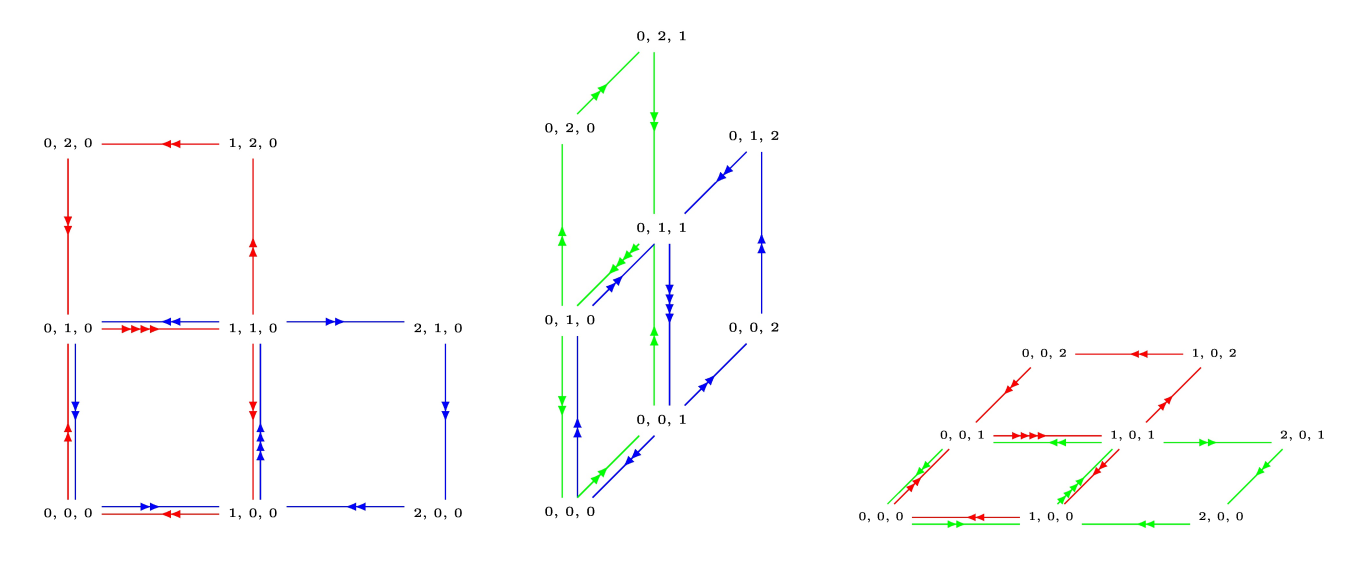}}
	\caption{Elementary excitations in SY loops: (a) the first excited states with $15$ units of energy and degeneracy 6; 
	(b) the second excited states with $32$ units of energy and degeneracy 6.}
    \label{SYLoopExcitations}
\end{figure}

In the previous section, we saw that Kleinert's dual theory for dislocations consists of symmetric loops $ \bar{\eta}_{ij}$, obeying a 
continuity condition. If the symmetry condition were relaxed ($\bar{\eta}_{ij} \neq  \bar{\eta}_{ji}$) then would have three 
equations, one for each color. On a regular square lattice, these loops will appear as simple plaquettes as shown in 
Fig.\ref{NSLoopExcitations}(a)-(b). For a given color, such loops may be realized as objects which are dual to the XY model in 
three dimensions\cite{Villain}. Therefore, for three colors with no interactions, we will have three independent XY models showing 
the same critical behaviour. The situation is quite different when there are interactions among the different colors. It is observed 
that the specific heat exponent $\alpha$ for the NS loop model with interactions is different from the non-interacting case 
(see Sec. IV). Contrary to the NS loops, the SY loops undergo a first-order transition \cite{Kleinert1,JankeKleinert}. Both the loop 
models have the energy function as 
\begin{equation}\label{PartitionFunction1D}
E(\bar{\eta}) = \frac{1}{2} \sum_{\mathbf{x},i,j} \bar{\eta}_{ij}^{2}(\mathbf{x}).
\end{equation}
Interestingly, the symmetry constraint on the SY loops results in thermodynamic properties that are very different from the NS loops 
(see Sec. V). Thus, it is far from obvious that the nature of the phase transition will be manifested from the structure of the Hamiltonian 
alone. It is also important to examine the nature of the loops and the manner in which they occur in the system 
(see Fig.\ref{SYLoopExcitations} and Appendix).      

We began our studies by looking at the NS loop model first. We wanted to examine their thermodynamic properties as well 
as their geometrical properties in the vicinity of the phase transition. In the companion paper (GMDb24), we studied the 
\textsl{percolation} transition in this model and calculated the respective critical exponents. 
To do this study, we have introduced an artificial anisotropy among the different tensor components. In our modified loop model, 
the diagonal components $\bar{\eta}_{ii}$ and the off-diagonal components $\bar{\eta}_{ij}$ appear with different energies. 
Therefore, we will have certain preferred directions where some loops would percolate sooner than others. It is important 
to note that the presence of such anisotropy does not affect the universal properties of this model. Rather, different choices 
of parameters will result in different transition temperatures. In addition to the anisotropy factor, the diagonal tensor 
components of the NS loops are made to interact with one another (inter-color coupling) by means of a coupling parameter. 
Although the interactions among the off-diagonal components are acceptable, they will be ignored for the present purpose. 
We will see that this minimal model will give rise to a phase diagram which is interesting in its own right (see Sec.IV A). 
With these extra features, the loop Hamiltonian in Eq.\eqref{PartitionFunction1D} will have the following modified expression 
\begin{eqnarray}\label{LoopHamiltonian4}
H(\bar{\eta}_{ij}) &=& \sum_{\mathbf{x},i \neq j} \Big[ A\bar{\eta}_{ii}^{2}(\mathbf{x}) + 
B\Big(\bar{\eta}_{ij}^{2}(\mathbf{x}) + \bar{\eta}_{ji}^{2}(\mathbf{x})\Big) \\ \nonumber
&+& D\bar{\eta}_{ii}(\mathbf{x})\bar{\eta}_{jj}(\mathbf{x}) \Big].
\end{eqnarray}
Where the parameters $A = 0.5$, $B = 0.25$, and $D = \pm 0.1,-0.2,-0.3$, etc., are in units of $k_{B}T$. The Hamiltonian in 
Eq.\eqref{LoopHamiltonian4} forms the basis of our studies in the present work. Using this Hamiltonian, we have examined 
the thermodynamic properties of both the SY and the NS loop systems. In the following section, we will see how to generate 
the NS loops and then using the NS loops, we will see how to generate the SY loops.  

\section{Method}

Computer experiments form an integral part of this work. Therefore, it is imperative to understand the process by 
which a network of such loops is actually generated or updated. We demonstrate this using the NS loop system. 
To distinguish the NS loops from the SY loops, we will represent them by $\eta_{ij}$. Here, the index $i$ denotes 
the direction and $j$ denotes the color of a loop variable. As an example, $\eta_{32}$ is a blue colored loop along 
the $z$-direction. From the previous sections, we have learnt that these loop variables are integer-valued quantities. 
In the ground state, the lattice is devoid of any loops. Therefore, we can write  
\begin{equation}\label{GroundState}
\eta_{ij}(\mathbf{x}) = 0 \quad \forall i,j,\mathbf{x}. 
\end{equation}
With this as the initial condition, let us suppose that a red colored loop ($j = 1$) with anti-clockwise 
chirality and oriented along the XY plane spontaneously occurs at the point $\mathbf{x} = (0,0,0)$ 
(Fig.\ref{NSLoopExcitations}(b) (extreme left)). Such a loop will occupy four bonds between the lattice points $(0,0,0), 
(1,0,0), (1,1,0)$, and $(0,1,0)$, respectively. The link between the points $(0,0,0)$ and $(1,0,0)$ will be occupied 
by the loop variable $\eta_{11}(0,0,0) = +1$. It is $+1$ because the flow is in the positive $x$-direction. Similarly, 
the link between the lattice points $(1,0,0)$ and $(1,1,0)$ will be occupied by the loop variable $\eta_{21}(1,0,0) = +1$. 
For the lattice point $(0,1,0)$, there is an inward flow from $(1,1,0)$, hence $\eta_{11}(0,1,0) = -1$. Finally, 
the circuit is completed by a flow from $(0,1,0)$ to $(0,0,0)$ by $\eta_{21}(0,0,0) = -1$. 
At an arbitrary lattice point $\mathbf{x}$, such a loop will be described by the following set of update equations.  

\begin{eqnarray}\label{NSLcase123A}
\eta_{11}(\mathbf{x})            &\rightarrow& \eta_{11}(\mathbf{x}) + 1   \\ \nonumber
\eta_{21}(\mathbf{x})            &\rightarrow& \eta_{21}(\mathbf{x}) - 1   \\ \nonumber
\eta_{11}(\mathbf{x} + \hat{y})  &\rightarrow& \eta_{11}(\mathbf{x} + \hat{y}) - 1   \\ \nonumber
\eta_{21}(\mathbf{x} + \hat{x})  &\rightarrow& \eta_{21}(\mathbf{x} + \hat{x}) + 1.
\end{eqnarray}

For a loop with clockwise chirality, the $+1$'s will be replaced by $-1$'s and vice-versa. In the same manner, the update 
equations can be written down for the other elementary loops shown in Fig.\ref{NSLoopExcitations}(a),(b). 

To obtain the update equations for the SY loops, we must first find a relationship between $\bar{\eta}$ and $\eta$.
In order to do that, one can \textsl{symmetrize} the NS loops by writing down the following \textsl{non-local} relationship 
\begin{equation}\label{SYMRelationB}
\bar{\eta}_{ij}(\mathbf{x}) = \Delta_{m}[\epsilon_{mjl}\eta_{il}(\mathbf{x}) + \epsilon_{mil}\eta_{jl}(\mathbf{x})].
\end{equation}
Where, $\Delta_{m}f(\mathbf{x}) \equiv f(\mathbf{x} + \hat{m}) - f(\mathbf{x})$ and $\epsilon$ is the third order 
anti-symmetric tensor. It is easy to check that $\bar{\eta}_{ij} = \bar{\eta}_{ji}$, and since $\Delta_{i}\eta_{ij}(\mathbf{x}) = 0$ 
we will have $\Delta_{i}\bar{\eta}_{ij}(\mathbf{x}) = 0$. Using the relation given in Eq.\eqref{SYMRelationB}, we can write down the 
update equations for the SY loops (see Eq.\eqref{SYcase3} in Appendix).

Using these update equations and the expression for the loop energy in Eq.\eqref{LoopHamiltonian4}, we have performed computer 
simulations using the MC method \cite{Murthy,Binder}. In accordance with the algorithm, the evolution of the canonical ensemble is represented 
by a first-order Markov chain of events. If $\Delta E$ represents the change in the energy of the system during the transition from a state 
$a$ to state $b$, then the transition probability $W(a \rightarrow b) = \min(1,\exp(-\beta\Delta E))$.

The outcomes of these computer experiments are presented in the following sections.

\section{Results}

We have performed extensive computer simulations for the loop models in regular three dimensional lattices with \textsl{periodic boundary conditions}. 
Their thermodynamic properties are presented in this section.

\subsection{Low and high-temperature results of the SY loop model}

\begin{figure}
    \centering
    \subfigure[]{\includegraphics[width=0.47\textwidth]{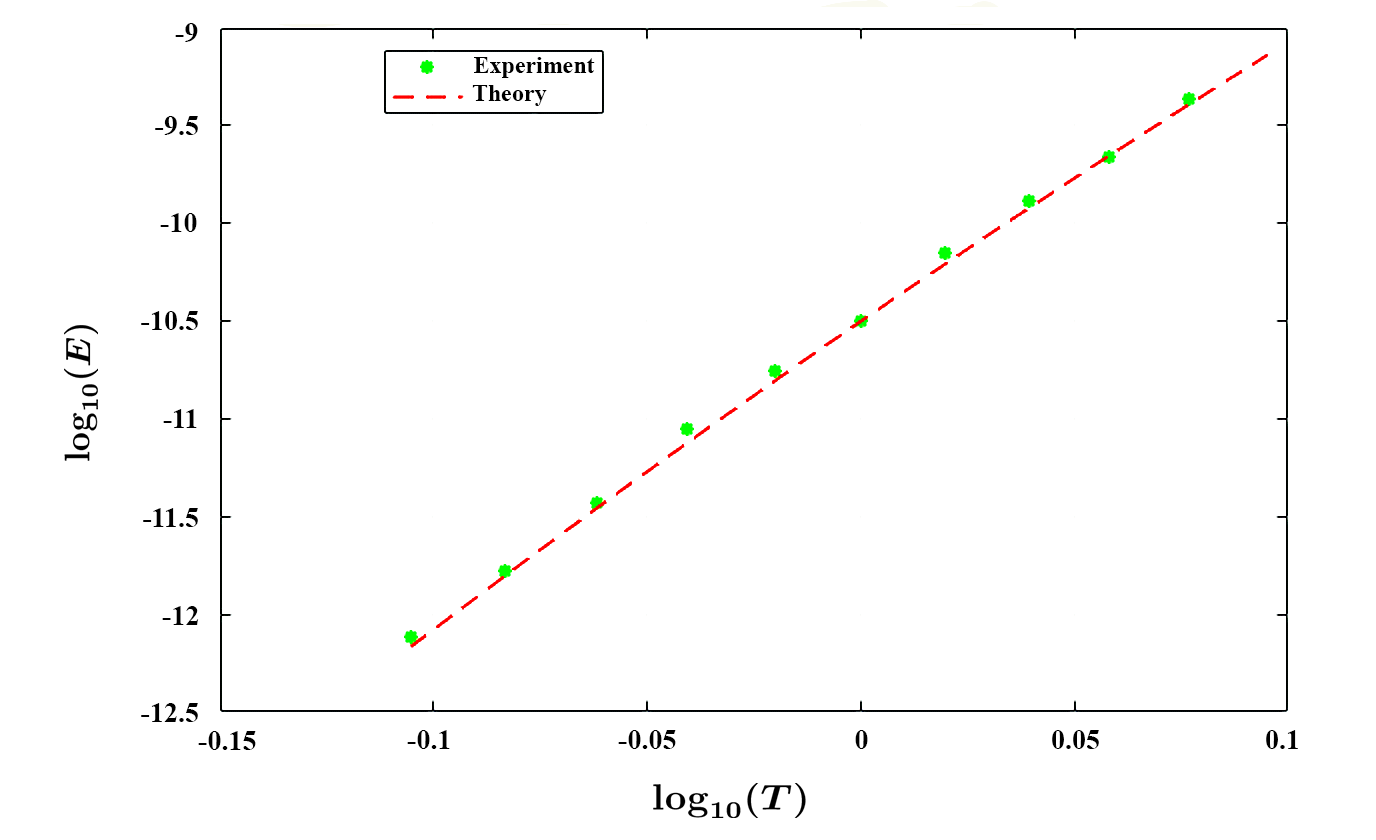}}
    \subfigure[]{\includegraphics[width=0.45\textwidth]{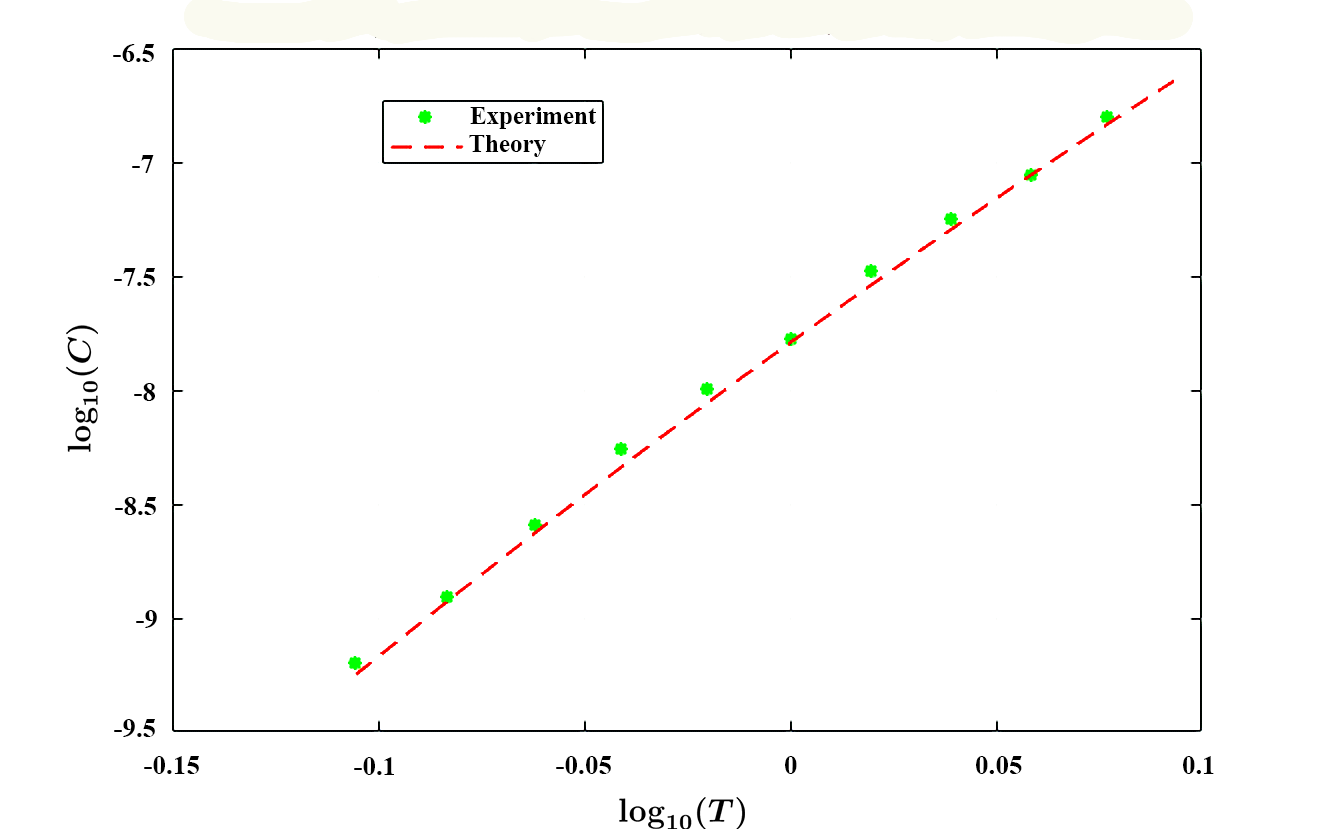}}
    \subfigure[]{\includegraphics[width=0.45\textwidth]{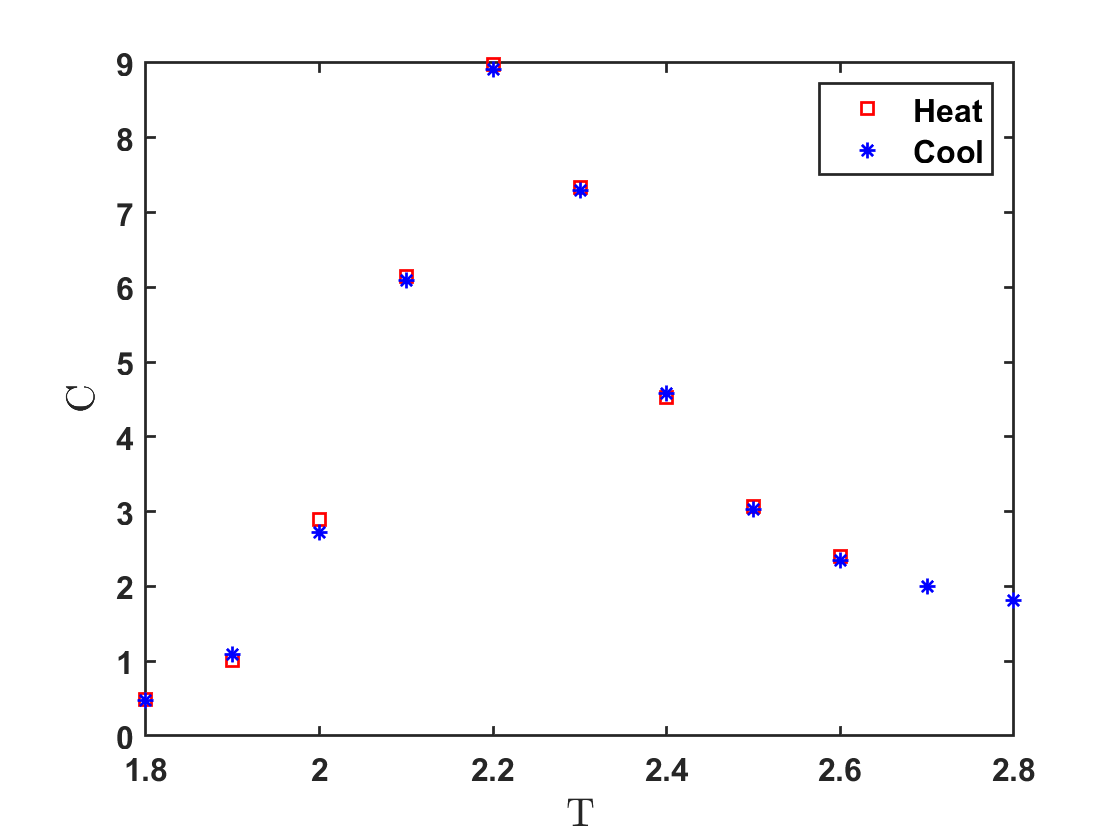}}
    \subfigure[]{\includegraphics[width=0.40\textwidth]{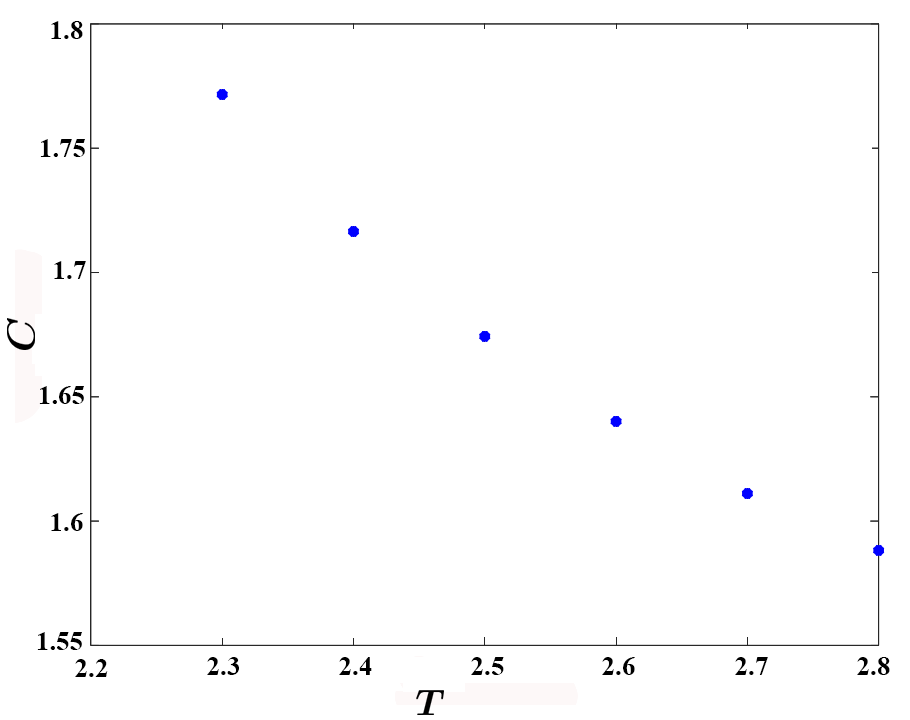}}
	\caption{Low and high-temperature results for the SY loops of size $L = 4$: (a) the $\log_{10}$-$\log_{10}$ plot for the internal energy ($E$) vs 
	temperature ($T$); (b) the $\log_{10}$-$\log_{10}$ plot for the specific heat ($C$) vs $T$; (c) the $C$ vs $T$ plot showing the transition; 
	(d) the $C$ vs $T$ plot in the high-temperature limit.}
    \label{SYMLoopsEC}
\end{figure}

In the previous section, we looked at the elementary excitations in the SY loop system. Using the expression for the energy in 
Eq.\eqref{LoopHamiltonian4} and the update equations in Appendix (Eq.\eqref{SYcase3}), we can calculate their energies. The most 
elementary excitations, shown in Fig.\ref{SYLoopExcitations}(a), carry $15$ units of energy each. The second excited state, shown in 
Fig.\ref{SYLoopExcitations}(b), carries $32$ units of energy. With the directions of the arrows reversed every excitation will have 
a multiplicity of $2$. Therefore, the total multiplicity of the first excited state is $3 \times 2 = 6$. The second excited state also 
has a total multiplicity of 6, which follows from the same argument as the first one. Using these information, we can write down the 
low-temperature partition function for the SY loops as 

\begin{equation}\label{SYloopPF1}
Z = 1 + 6\exp(-15\beta) + 6\exp(-32\beta).
\end{equation}

For a canonical ensemble, we know that all the thermodynamic quantities can be calculated from the partition function. For example, 
the free energy $F$ for the loop system is proportional to $\ln(Z)$ in Eq.\eqref{SYloopPF1}. If $x \equiv 6\exp(-15\beta) + 6\exp(-32\beta)$, 
then in the low-temperature limit, $\ln(1+x) \approx x$. Therefore, the free energy of the loops in this limit is given by   

\begin{equation}\label{SYloopPF2}
F = -k_{B}T\ln(Z) = 6\exp(-15\beta) + 6\exp(-32\beta).
\end{equation}

The internal energy $E$ can be obtained from Eq.\eqref{SYloopPF2} by taking the first derivative with respect to $\beta$ as 

\begin{equation}\label{SYloopPF3}
E = 90 \exp(-15\beta) + 192 \exp(-32\beta). 
\end{equation}

Similarly, the specific heat $C$ can be obtained by taking the second derivative of Eq.\eqref{SYloopPF3} with respect to $T$ as   

\begin{equation}\label{SYloopPF4}
C = \beta^{2}[1350 \exp(-15\beta) + 6144 \exp(-32\beta)].
\end{equation}

In the low-temperature limit, the acceptance rate is low due to the high energy of the SY loops. We have performed the simulations for 
temperatures not too far from the transition temperature ($T < 2.14$). The internal energy and the specific heat are plotted in the $\log_{10}$ 
scale for good comparison (Fig.\ref{SYMLoopsEC}). The results obtained from the simulation data are found to be in good agreement with 
the theoretical predictions.

The order-disorder transition in SY loops is characterized by a peak in the specific heat capacity at T = 2.2 (Fig.\ref{SYMLoopsEC}(c)).
At high-temperatures, the system is in a disordered state. If $N$ be the total number of lattice sites, then each site can have 
$6$ loops ($3$ diagonal and $3$ off-diagonal elements). Nevertheless, we also have $3$ continuity equations, which act as constraints. 
Therefore, the total number of degrees of freedom is reduced to $(6-3)N = 3N$. Due to the quadratic nature of the Hamiltonian, the 
internal energy associated with each degree of freedom will be $k_{B}T/2$ (Equipartition Theorem) \cite{LandauLifshitz,Pathria}. 
Therefore, the specific heat capacity per unit volume at high-temperatures will assume a value given by $3k_{B}/2$ (Fig.\ref{SYMLoopsEC}(d)). 
In the next section, we will look at the metastable behaviour of these loops when they discontinuously transit into the disordered state.

\subsection{Metastability in SY loop model}
\begin{figure}
    \centering
    \subfigure[]{\includegraphics[width=0.328\textwidth]{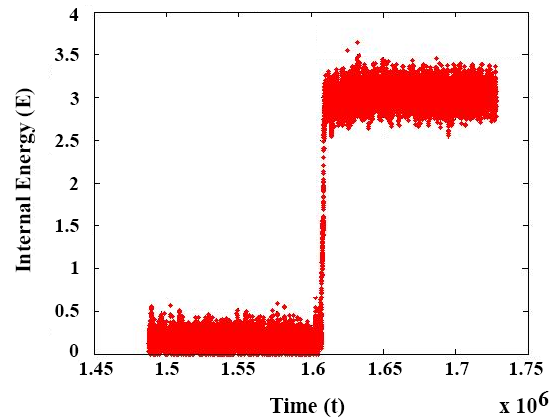}}
    \subfigure[]{\includegraphics[width=0.328\textwidth]{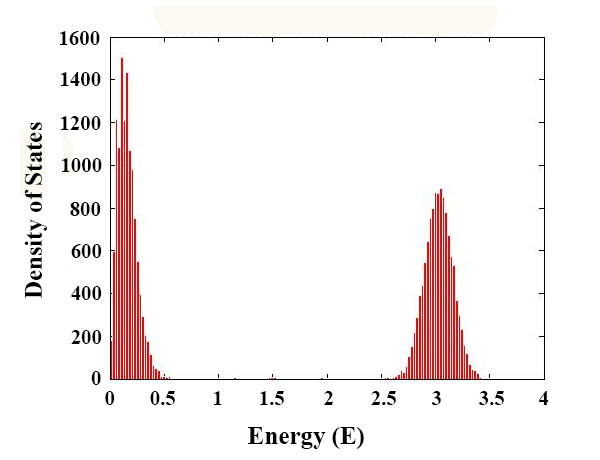}}
    \subfigure[]{\includegraphics[width=0.328\textwidth]{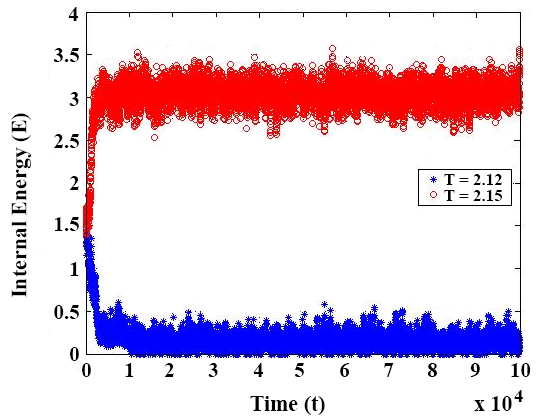}}
    \subfigure[]{\includegraphics[width=0.298\textwidth]{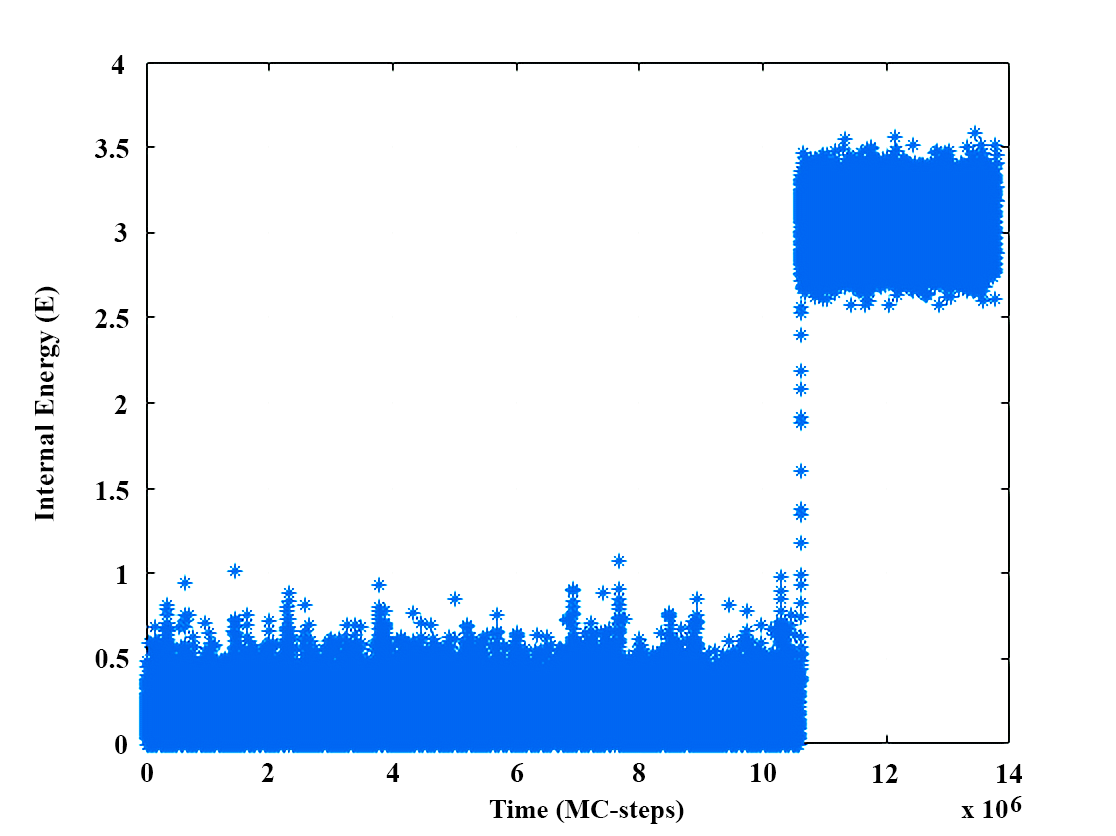}}
    \subfigure[]{\includegraphics[width=0.298\textwidth]{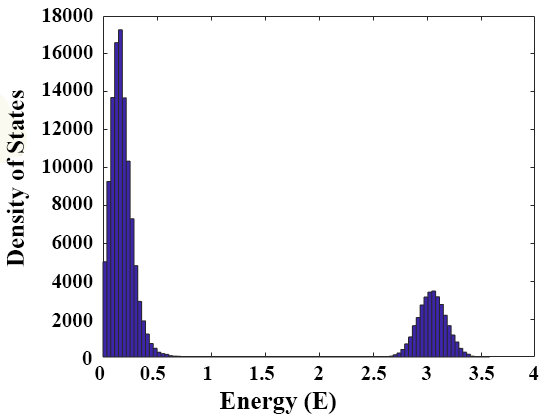}}
    \subfigure[]{\includegraphics[width=0.338\textwidth]{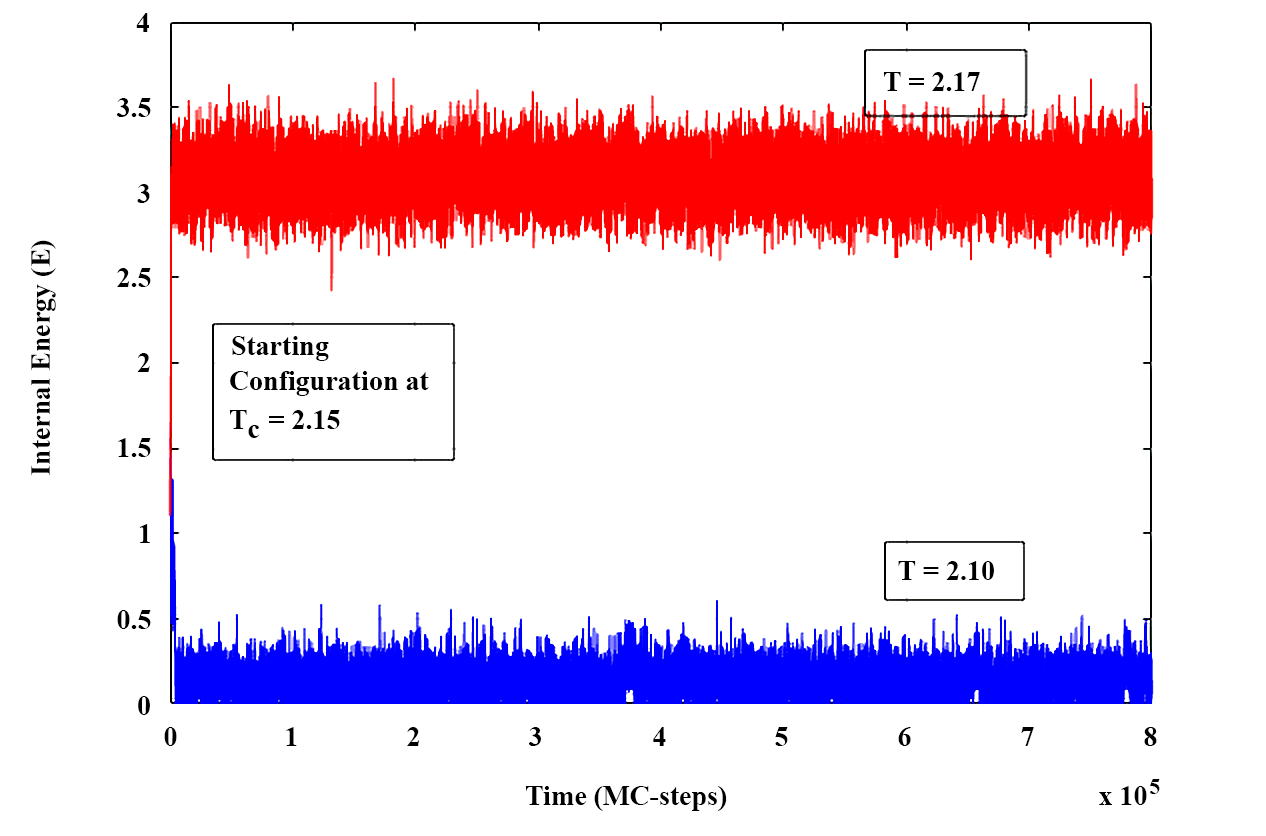}}
	\caption{Metastable behaviour seen in SY loops of size $L = 8$, at temperatures close to the critical point, $T = 2.14,2.15$: 
	(a) $E$ vs time (MC-steps) plot for case 1; (b) density of states vs $E$ plot for case 1; (c) $E$ vs time plot for the system 
	at $T > T_{c}$ (red) and $T < T_{c}$ (blue), for case 1; (d) $E$ vs time plot for case 2; (e) density of states vs $E$ plot for case 1; 
	(f) $E$ vs time plot for the system at $T > T_{c}$ (red) and $T < T_{c}$ (blue), for case 2.}
    \label{EvolutionSymm}
\end{figure}

To study the metastability in SY loops, we took a system of size $L = 8$. The system was made to evolve from a low-temperature phase to a high 
temperature phase. The internal energy as a function of time shows a jump discontinuity, while the system makes a transition into the disordered 
state Fig.\ref{EvolutionSymm}(a),(d). We found the transition temperature to be within the range of $ 2.14 < T_{c} < 2.15$. We did a histogram plot 
for the density of states and found two distinct humps at two different values of the internal energy (Fig.\ref{EvolutionSymm}(b),(e)). 
Within the temperature range of $2.14 < T < 2.15$, we ran the simulations and stored a few loop configurations. These loop configurations 
were then used as initial conditions to study the evolution of the system above and below the critical temperature $T_{c}$. 
Starting from a mixed-state configuration, the system transited to the high-temperature phase for temperature $T > T_{c}$. While for 
$T < T_{c}$, the system was found to transit into the low-temperature phase (Fig.\ref{EvolutionSymm}(a),(d)). The latent 
heat of melting at $T_{c}$ is simply the change in the internal energy $\Delta E \approx 2.75$ (Fig.\ref{EvolutionSymm}(f)). 
Using the relation
\begin{equation}\label{InternalEnergy_Entropy}
\Delta E = T_{m}\Delta S,
\end{equation}
we calculated the entropy change $\Delta S = 1.282 \pm 0.003$. Our findings qualitatively agree with the results of Janke and 
Kleinert\cite{JankeKleinert}.


\subsection{Low and high-temperature results of the NS loop model}

\begin{figure}
    \centering
    \subfigure[]{\includegraphics[width=0.46\textwidth]{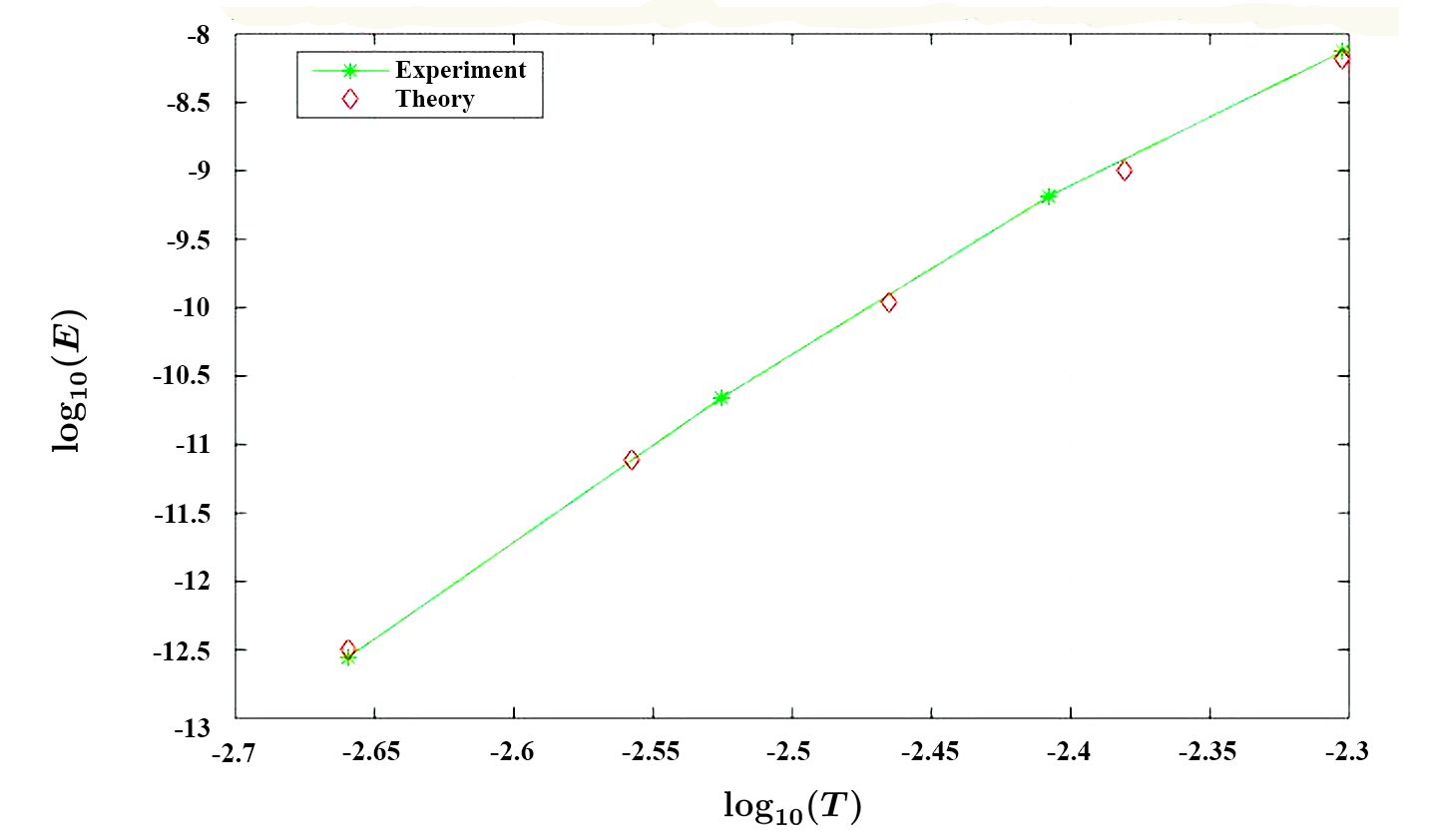}}
    \subfigure[]{\includegraphics[width=0.49\textwidth]{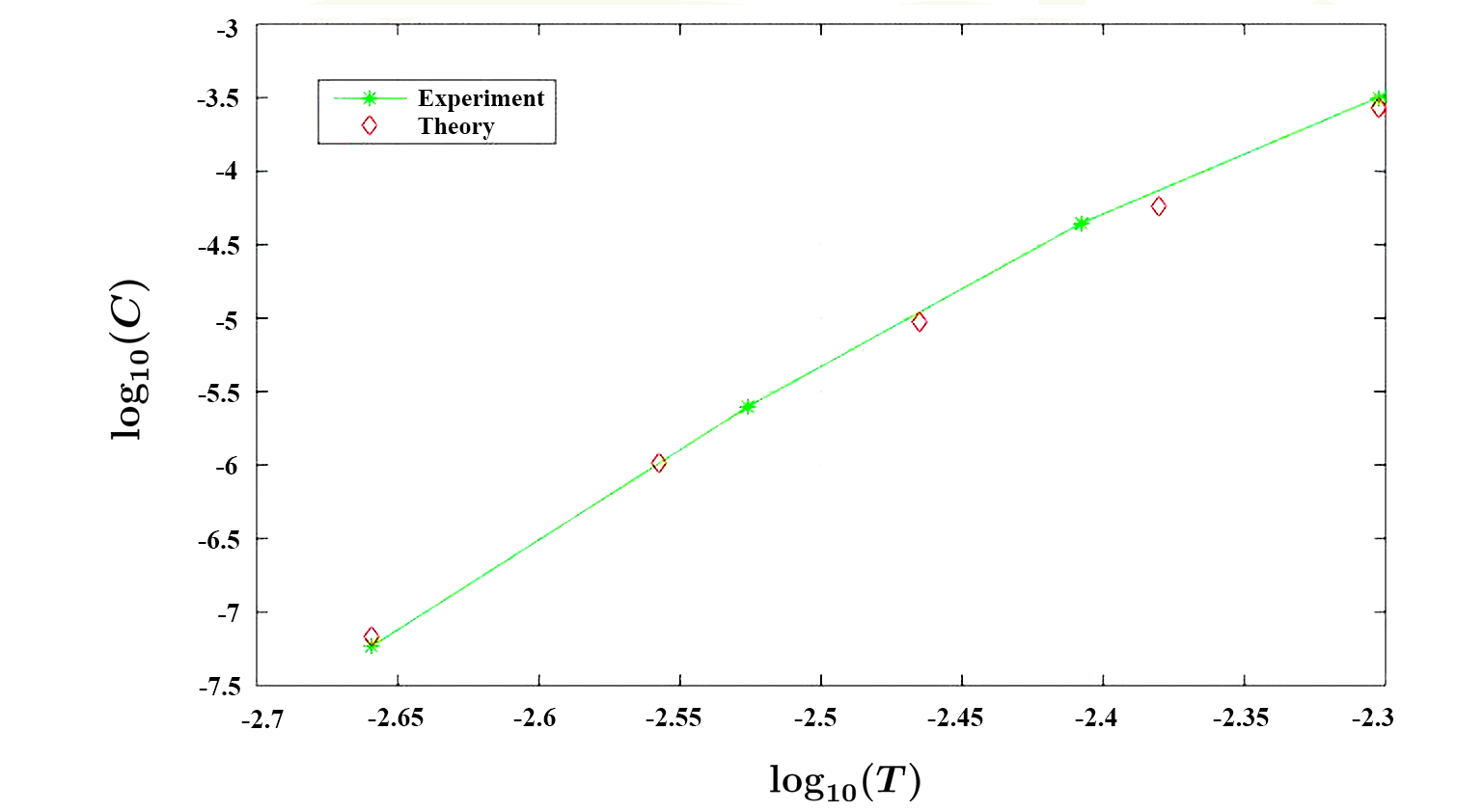}}
	\caption{Low temperature results for the NS loops of size $L = 4$: (a) $\log_{10}$-$\log_{10}$ plot for the $E$ vs $T$; 
	(b) $\log_{10}$-$\log_{10}$ plot for the $C$ vs $T$.} 
    \label{NSLoopsEC}
\end{figure}

We have seen in the previous section, that the elementary excitations in the NS loop system are simpler than their 
symmetric cousins. The energies of these excitations can be calculated using Eq.\eqref{LoopHamiltonian4}. The first 
excited state in Fig.\ref{NSLoopExcitations}(a) has $1$ unit of energy. Each loop can occur with a multiplicity of $2$ 
(clockwise and anti-clockwise). Therefore, the first excited state has a total multiplicity of $3 \times 2 = 6$. The second 
excited state in Fig.\ref{NSLoopExcitations}(b) has $1.5$ units of energy. In this case, every color occurs twice with 
both chiralities. Therefore, the total multiplicity is $2 \times 2 \times 3 = 12$. Similarly, the third excited state 
has $2$ units of energy and a total multiplicity of $32$ ($16$ orientations and two chiralities, Fig.\ref{NSLoopExcitations}(a),(b)). 
The fourth excited state has $2.5$ units of energy with a total multiplicity of $12$ ($6$ orientations and two chiralities, 
Fig.\ref{NSLoopExcitations}(c) ). With the first two excited states, we can write down the approximate low-temperature partition 
function, which has the following form

\begin{equation}\label{NSLoopsLowT1A}
Z = 1 + 6\exp(-1.0\beta) + 12\exp(-1.5\beta) 
\end{equation}

Using Eq.\eqref{NSLoopsLowT1A}, the internal energy and the specific heat will be

\begin{eqnarray}\label{NSLoopsLowT2B}
E &=& 6\exp(-1.0\beta) + 18\exp(-1.5\beta),        \\    \nonumber
\textrm{and} \quad C &=& \beta^{2}[6\exp(-1.0\beta) + 27\exp(-1.5\beta)].
\end{eqnarray}

The transition temperature for the NS loops is $T_{c} = 0.212$ (see TABLE.\ref{SpecificHeat_exponentsD}). So, we had to go 
to much lower temperatures to perform the simulations. To get a good comparison, we plotted the internal energy and the specific 
heat capacity in the $\log_{10}$ scale (Fig.\ref{NSLoopsEC}). We find that the simulation data and the analytical expressions 
in Eq.\eqref{NSLoopsLowT2B} agree with one another.

\begin{figure}
    \centering
    \subfigure[]{\includegraphics[width=0.328\textwidth]{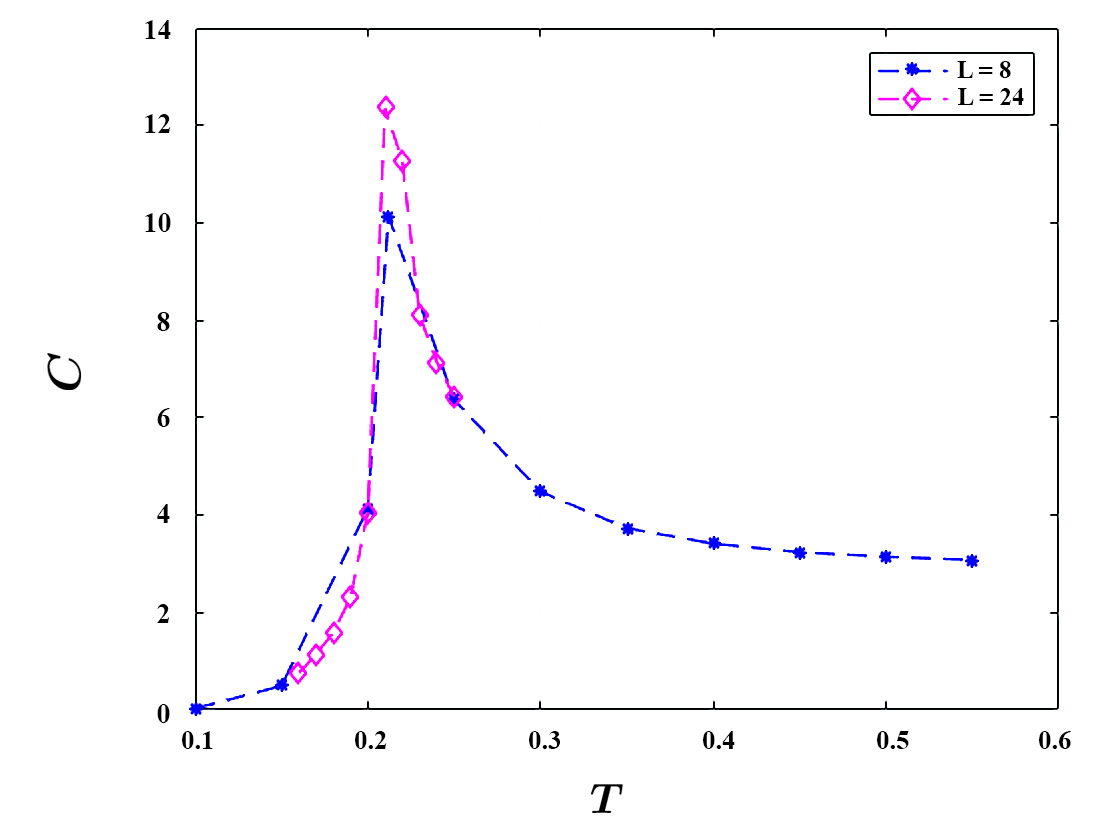}}
    \subfigure[]{\includegraphics[width=0.328\textwidth]{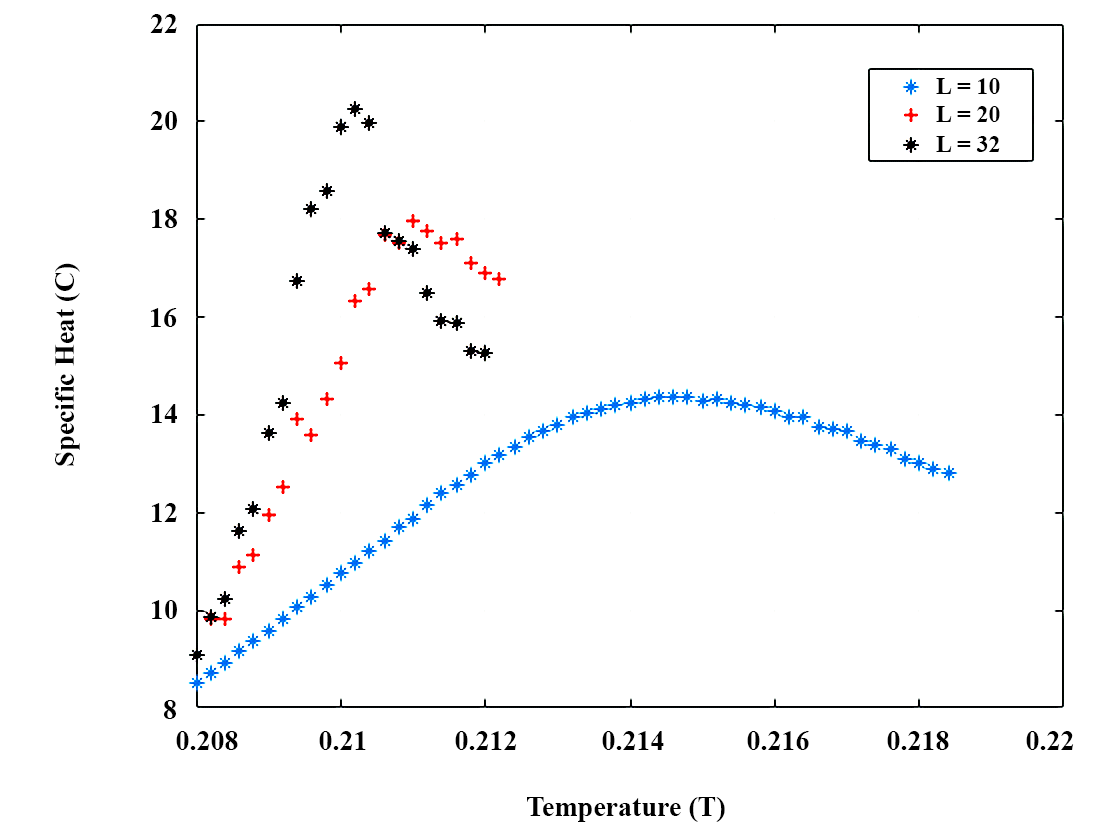}}
    \subfigure[]{\includegraphics[width=0.328\textwidth]{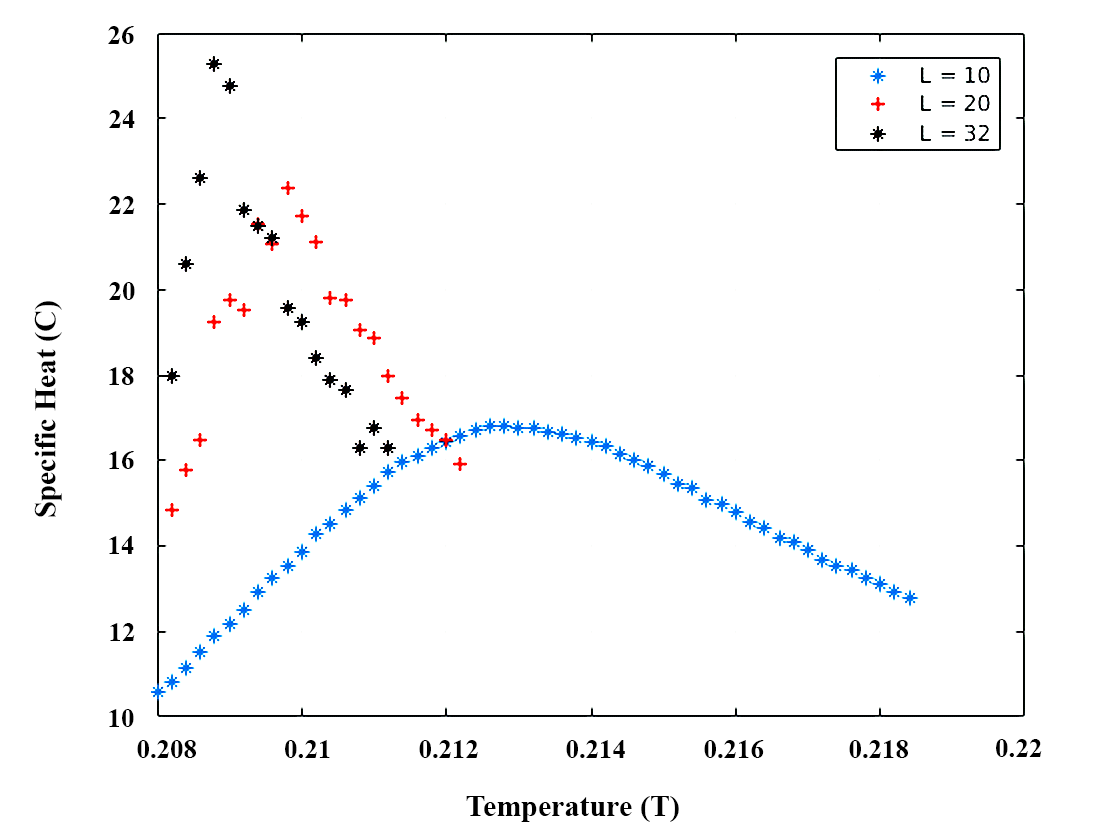}}
	\caption{Finite temperature results for NS loops with varying inter-color coupling $D$: (a) $C$ vs $T$ plot for $D = 0.1$ with sizes $L = 8, 24$; 
	(b) $C$ vs $T$ plot for $D = -0.2$ and system sizes $L = 10, 20, 32$; (c) $C$ vs $T$ plot for $D = -0.3$ with sizes $L = 10,20,32$.}
    \label{NSLoopsCTD}
\end{figure}

In the high-temperature phase, the NS loops are in a completely disordered state. In this state, each lattice site 
can have $9$ degrees of freedom: $3$ diagonal and $6$ off-diagonal elements. The continuity condition will reduce 
it to $9 - 3 = 6$ degrees of freedom. Therefore, at large $T$ the specific heat per unit volume will be equal $3k_{B}T$ 
(Fig.\ref{NSLoopsCTD}(a)). 

As the NS loops continuously go to the high-temperature phase, it passes through a critical region. During this transition, 
the internal energy smoothly goes to a finite value and the specific heat capacity attains a peak value. However, the peak 
value of the specific heat capacity has a system size dependence. This suggests that the peak value, at the critical temperature, 
must diverge for an infinitely large system. Therefore, we must be able to locate the critical temperature by doing a finite 
size scaling analysis. In the next section, we will calculate the specific heat exponent for the NS loop system. 
We will also see how this exponent varies with the inter-color coupling parameter $D$.

\subsection{Finite-size scaling and the effects of large inter-Color couplings in NS loops}

\begin{figure}
    \centering
    \subfigure[]{\includegraphics[width=0.45\textwidth]{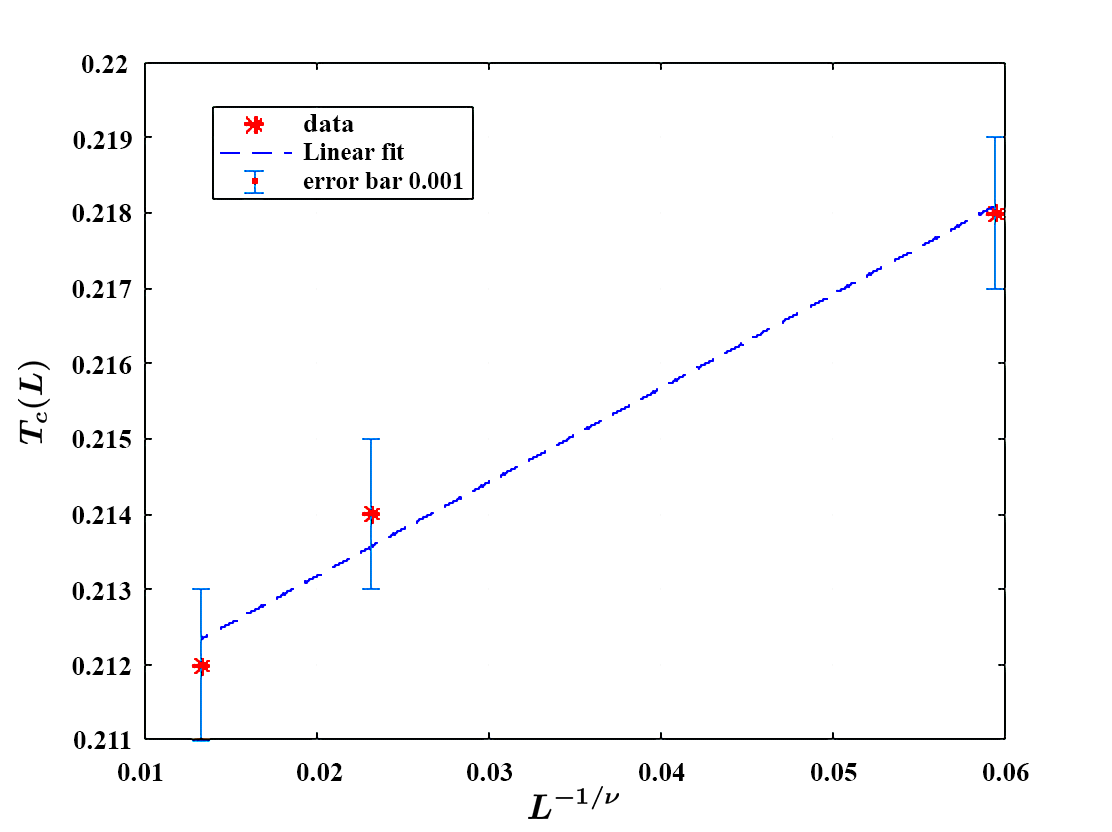}}
    \subfigure[]{\includegraphics[width=0.45\textwidth]{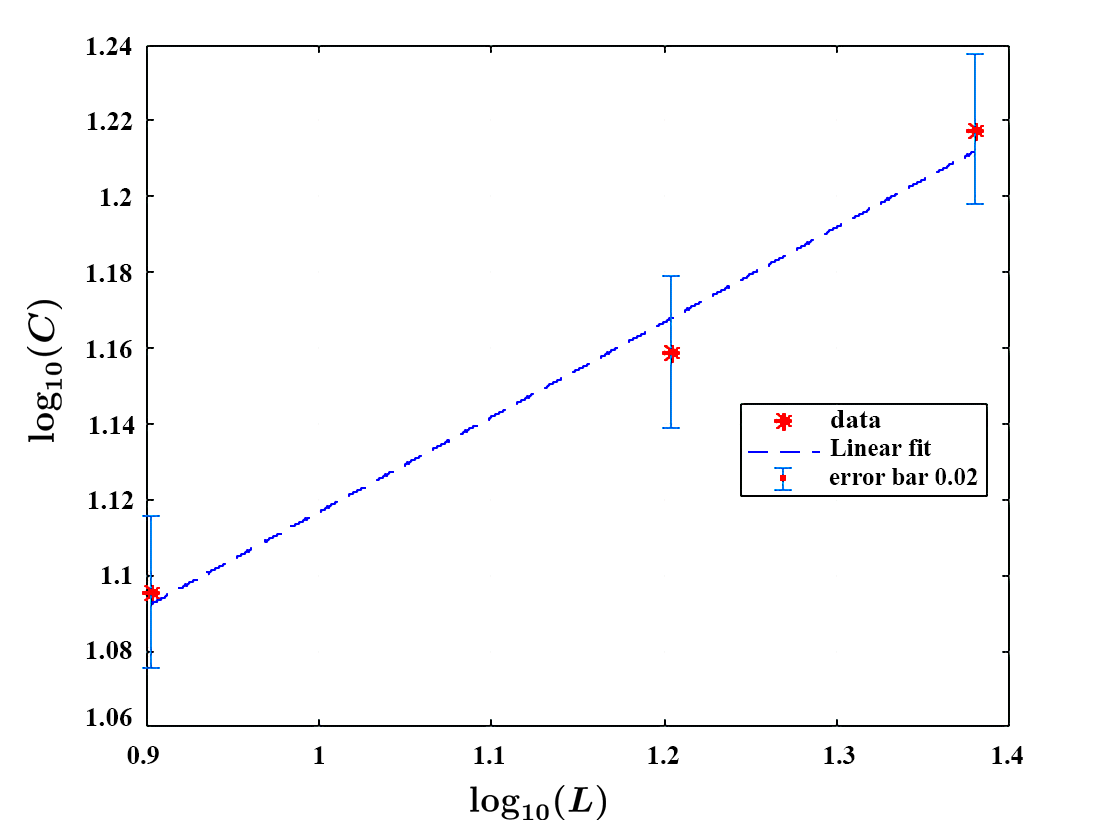}}
    \subfigure[]{\includegraphics[width=0.45\textwidth]{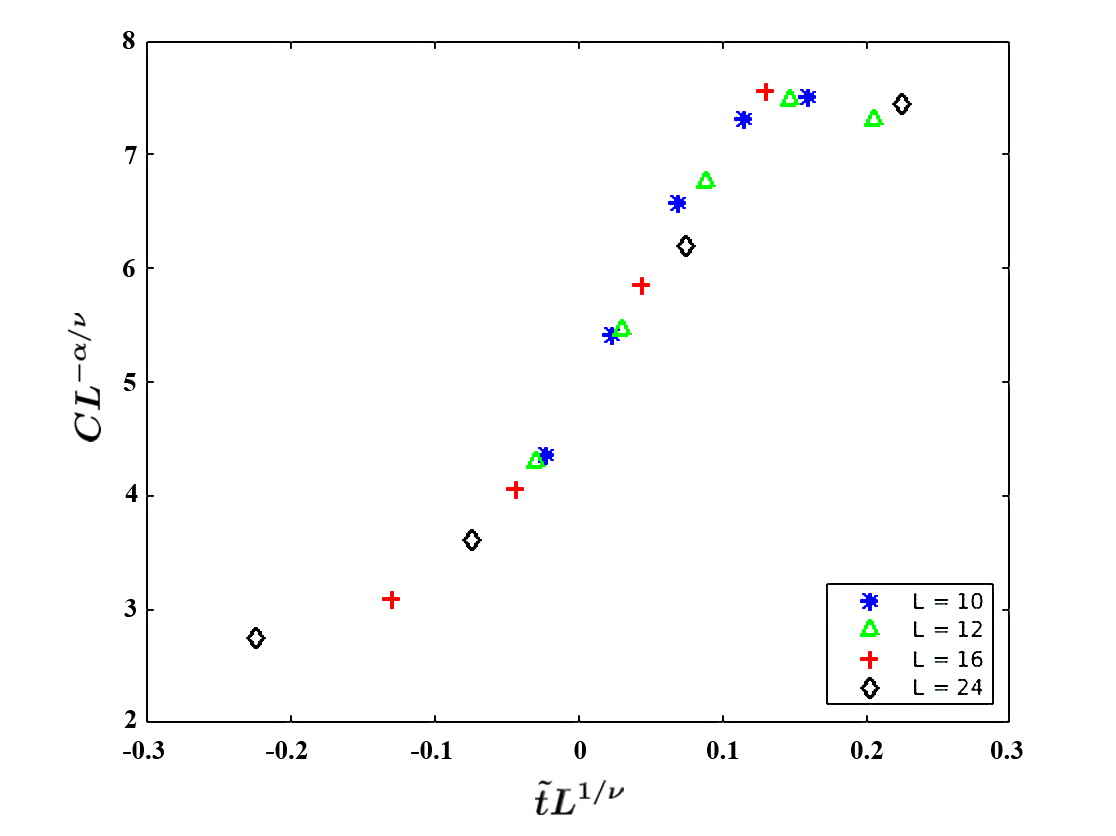}}
    \subfigure[]{\includegraphics[width=0.45\textwidth]{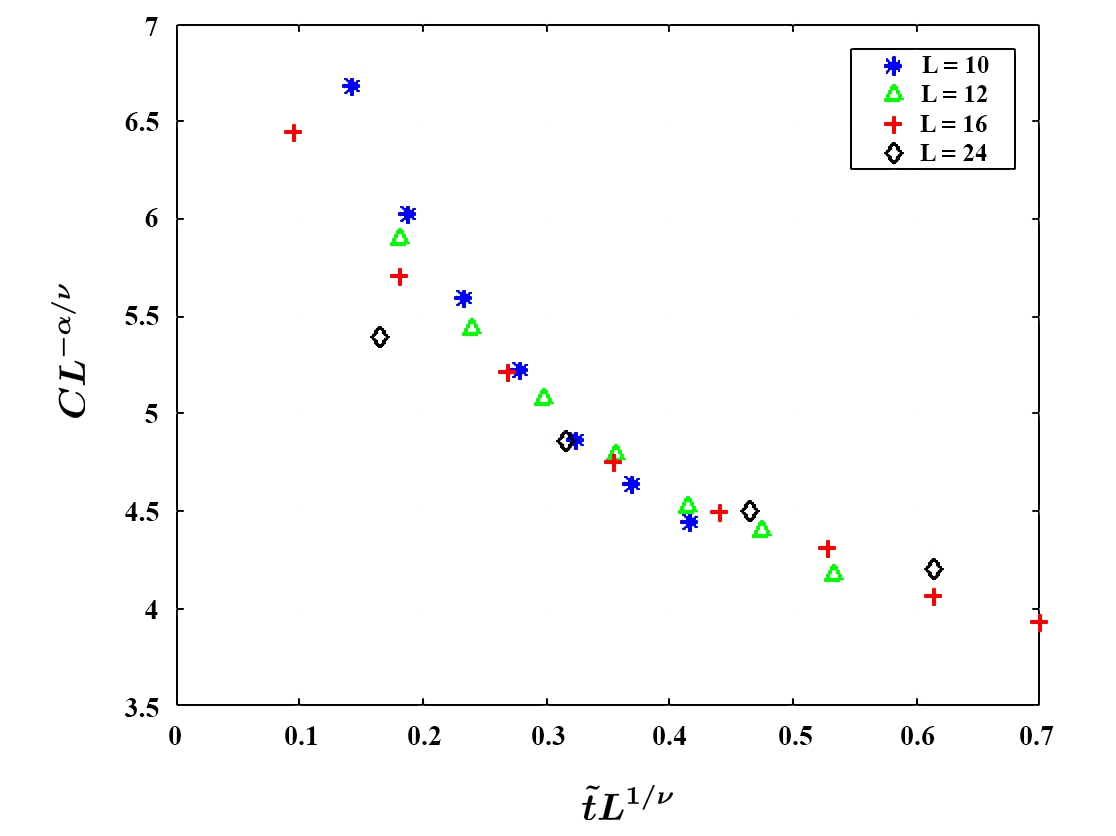}}
	\caption{Finite-size scaling results for NS loops with $D = 0.1$: (a) $T_{c}$ vs $L^{-1/\nu}$ plot for the correlation 
	length exponent $1/\nu = 1.3(0058)$; (b) $\log_{10}$-$\log_{10}$ plot for $C$ vs $L$ data giving the exponent $\alpha/\nu = 0.25(0008)$; 
	(c) the $CL^{-\alpha/\nu}$ vs $\tilde{t}L^{1/\nu}$ plot for $T < T_{c}$; (d) $CL^{-\alpha/\nu}$ vs $\tilde{t}L^{1/\nu}$ plot for 
	$T > T_{c}$ ($\tilde{t} = \vert T - T_{c}\vert$).}
    \label{NSLoopsFSSa}
\end{figure}

During the transition in NS loops, the correlation length $\xi \rightarrow \infty$, as $L \rightarrow \infty$. For a finite-sized system, 
the correlation length is limited by the system size ($\xi \sim L$). According to the scaling theory of second-order phase transitions, the 
specific heat capacity $C$ has a scaling relationship with the reduced temperature $\tilde{t} = \vert T - T_{c}\vert $, which reads as follows 
\begin{equation}
C_{\infty}(L) = L^{\frac{\alpha}{\nu}} \bar{C}_{\infty}(\tilde{t} L^{\frac{1}{\nu}}).
\end{equation}

\begin{figure}
    \centering
    \subfigure[]{\includegraphics[width=0.45\textwidth]{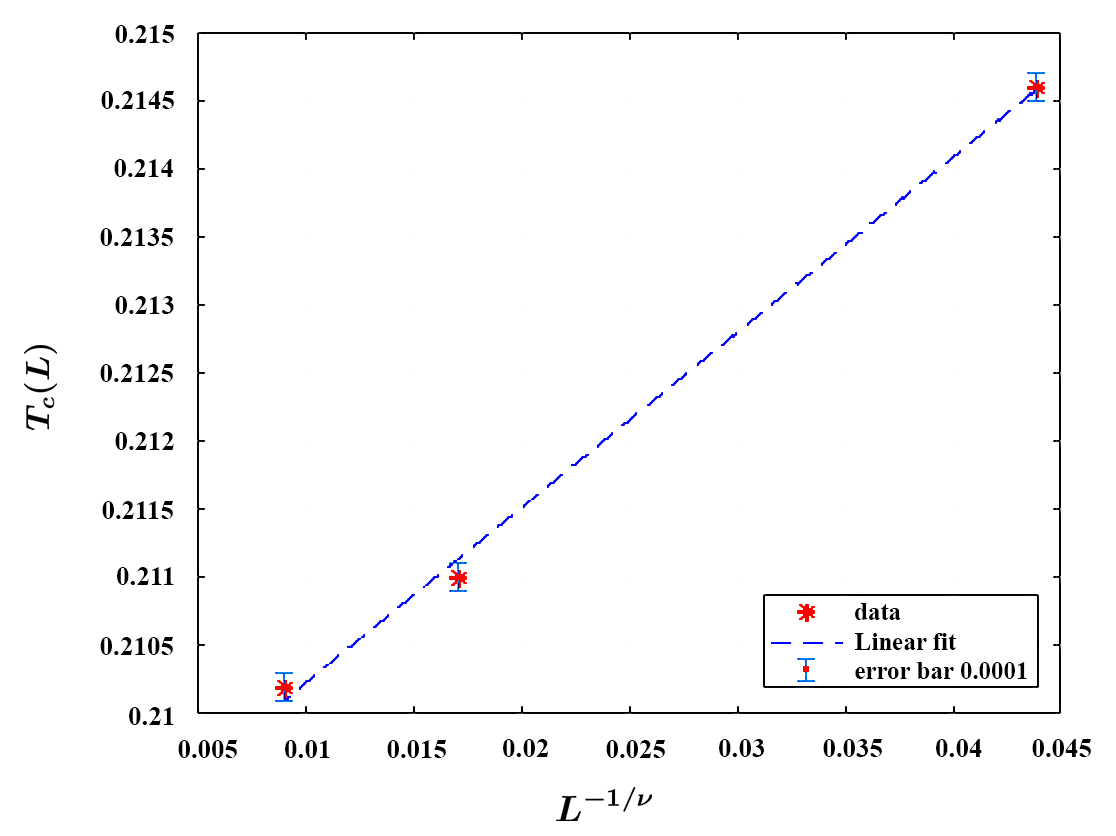}}
    \subfigure[]{\includegraphics[width=0.45\textwidth]{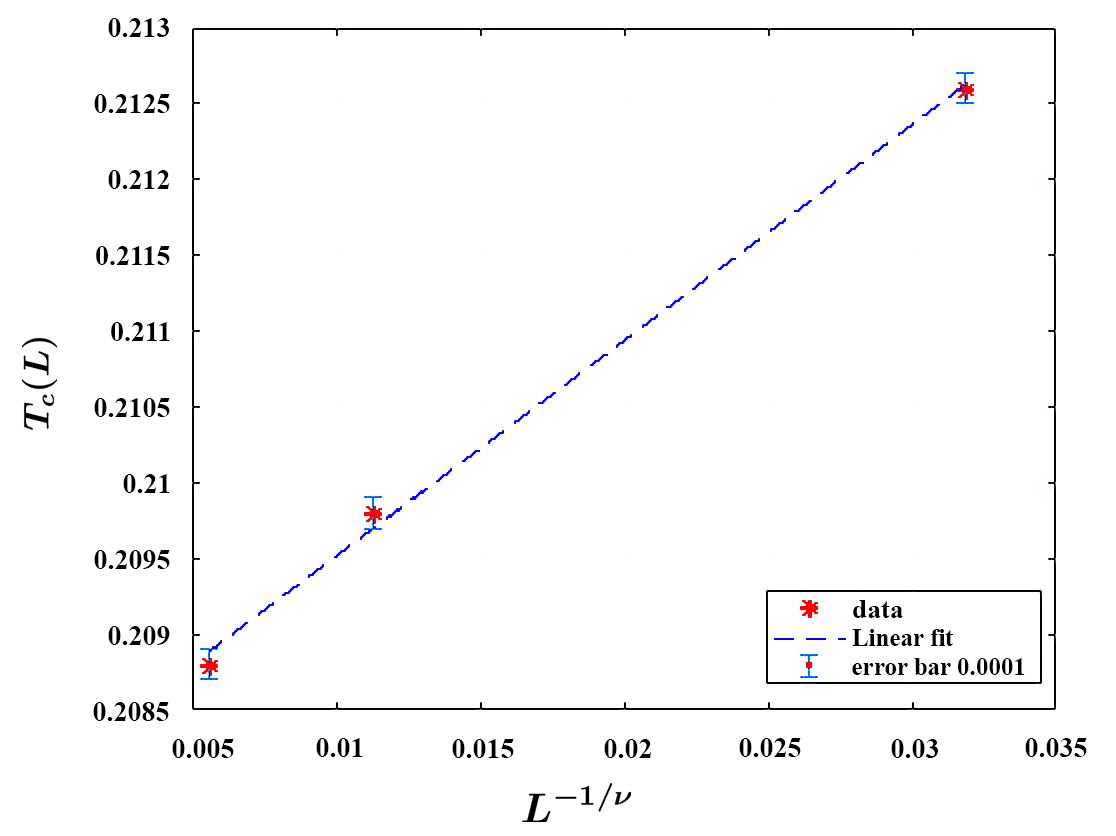}}
    \subfigure[]{\includegraphics[width=0.45\textwidth]{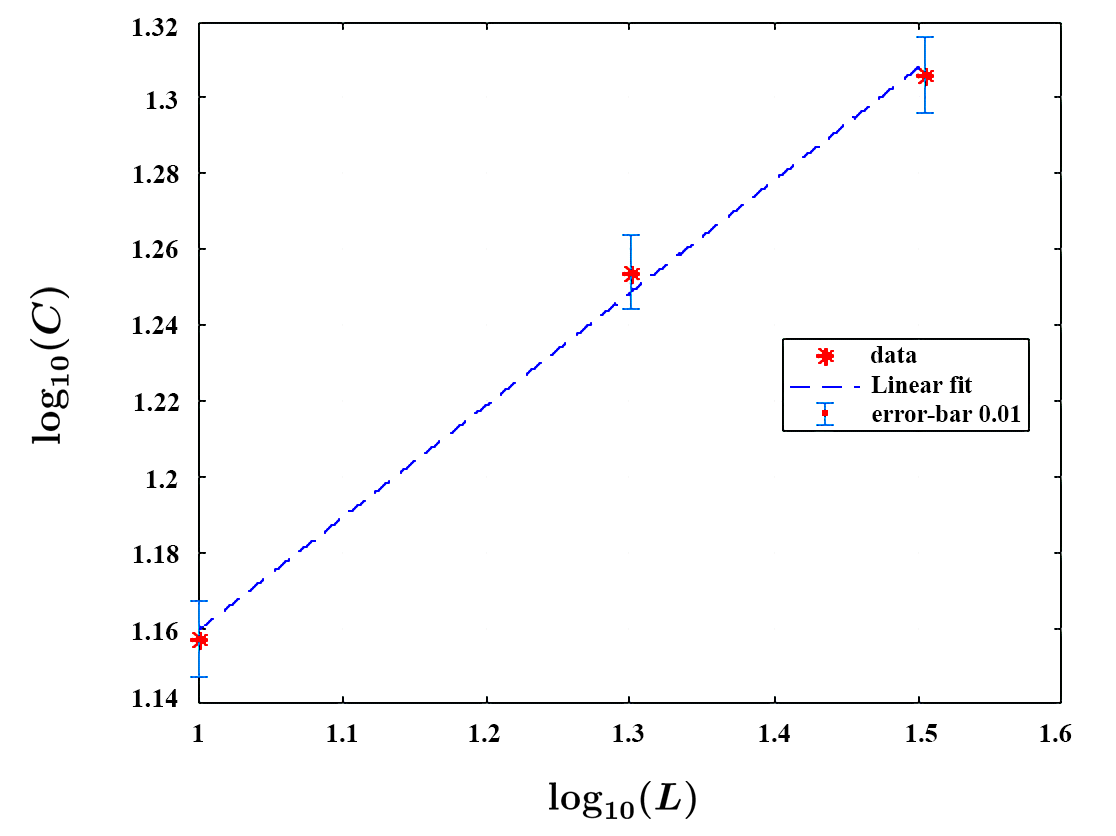}}
    \subfigure[]{\includegraphics[width=0.45\textwidth]{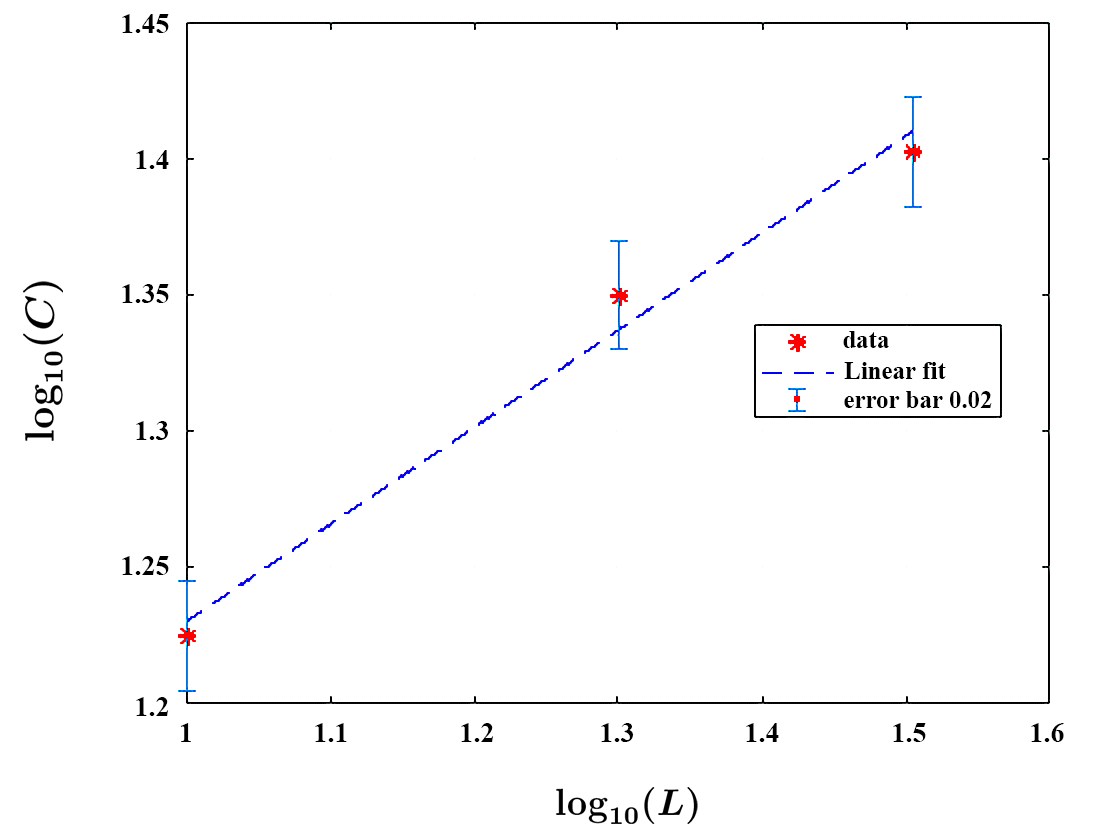}}
    \subfigure[]{\includegraphics[width=0.45\textwidth]{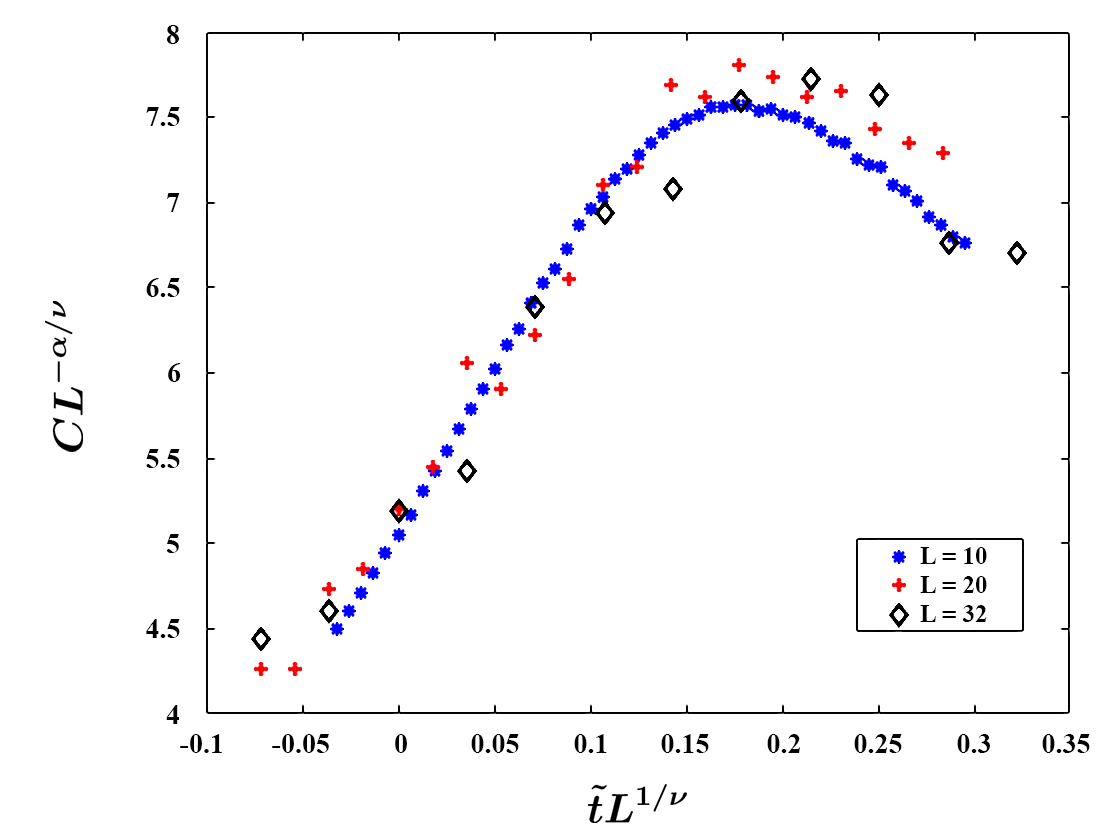}}
    \subfigure[]{\includegraphics[width=0.45\textwidth]{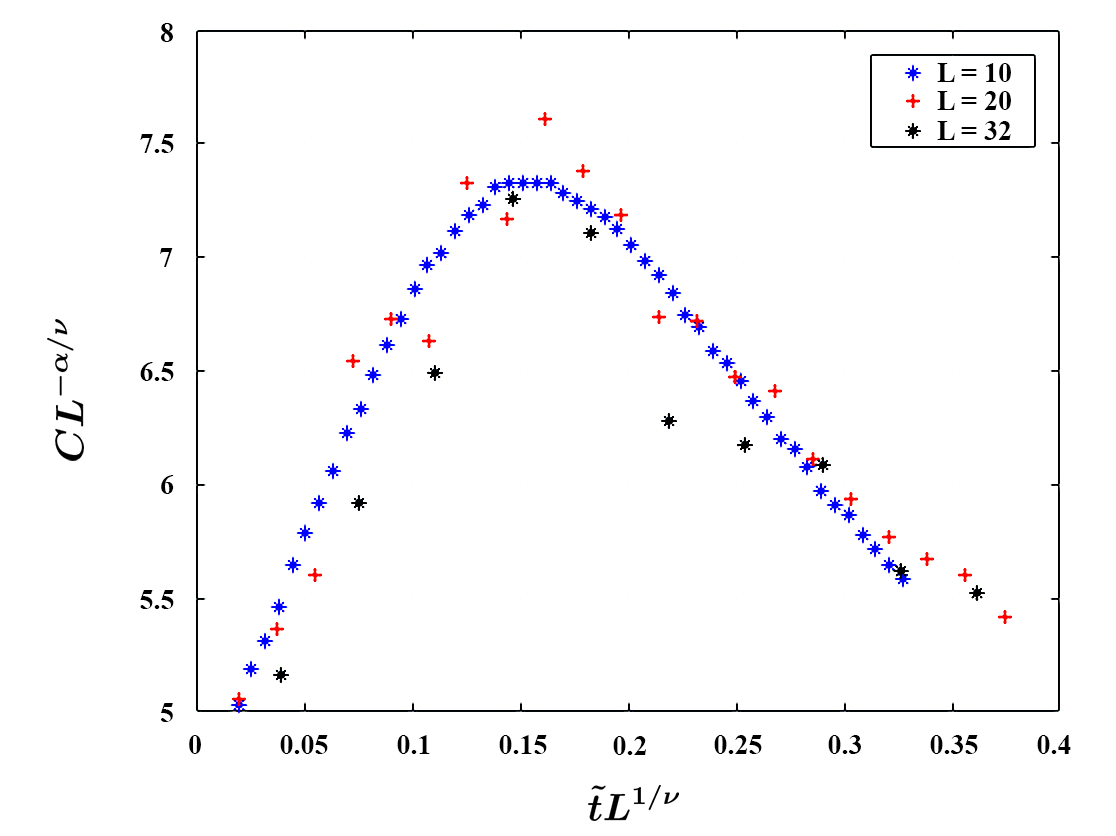}}
\caption{Finite-size scaling results for NS loops with $D = -0.2,-0.3$: (a) the $T_{c}$ vs $L^{-1/\nu}$ plot for $1/\nu = 1.3(0023)$, for case 1; 
(b) the $T_{c}$ vs $L^{-1/\nu}$ plot for $1/\nu = 1.3(0023)$, for case 2; (c) the $\log_{10}$-$\log_{10}$ plot for $C$ vs $L$ data for $\alpha/\nu = 0.27(0008)$, 
for case 1; (d) the $\log_{10}$-$\log_{10}$ plot for $C$ vs $L$ data for $\alpha/\nu = 0.34(008)$, for case 2; (e) the $CL^{-\alpha/\nu}$ vs $\tilde{t}L^{1/\nu}$ 
plot for case 1; (f) the $CL^{-\alpha/\nu}$ vs $\tilde{t}L^{1/\nu}$ plot for case 2 ($\tilde{t} = \vert T - T_{c}\vert$).}
    \label{NSLoopsFSSb}
\end{figure}

Where $\alpha$ is the specific heat exponent and $\nu$ is the correlation length exponent. We have performed simulations for systems of 
varying sizes. For different values of $D$, we have calculated $\alpha$, $\nu$, and $T_{c}$ and listed them in TABLE.\ref{SpecificHeat_exponentsD}. 
Using these values, we have obtained the data collapse for the specific heat vs temperature curves 
(Fig.\ref{NSLoopsFSSb}(e),(f)). 

\begin{center}
\begin{table}
 \begin{tabular}{|| c c c c ||}
 \hline
  $D$  & $T_{c}$ &  $1/\nu$  & $\alpha/\nu$ \\ [0.5ex]
 \hline\hline
  0.1  &  0.212 &  1.3(058) & 0.25(001)      \\
 -0.2  &  0.209 &  1.3(023) & 0.27(008)      \\
 -0.3  &  0.208 &  1.3(023) & 0.34(008)      \\
 \hline
\end{tabular}
\caption{Comparison of the critical temperatures ($T_{c}$), correlation exponent ($\nu$), and specific heat exponent ($\alpha$)
	for different inter-color couplings ($D$).}
\label{SpecificHeat_exponentsD}
\end{table}
\end{center}

We find that the inter-color coupling has a profound effect on the nature of the transition in NS loops. Finite-size scaling
studies for large negative values of $D$ (i.e., $D = -0.2,-0.3$) show a lowering of the transition temperature $T_{c}$ and an 
increase in the specific heat exponent $\alpha/\nu$ (TABLE.\ref{SpecificHeat_exponentsD}). We will see in the next section that 
for $D = -0.5$, the NS loops behave like the SY loops and undergo a first-order phase transition.


\subsection{First order transition in Non-Symmetric loops at `$D = -0.5$' and the tri-critical phase.}

\begin{figure}
    \centering
    \subfigure[]{\includegraphics[width=0.295\textwidth]{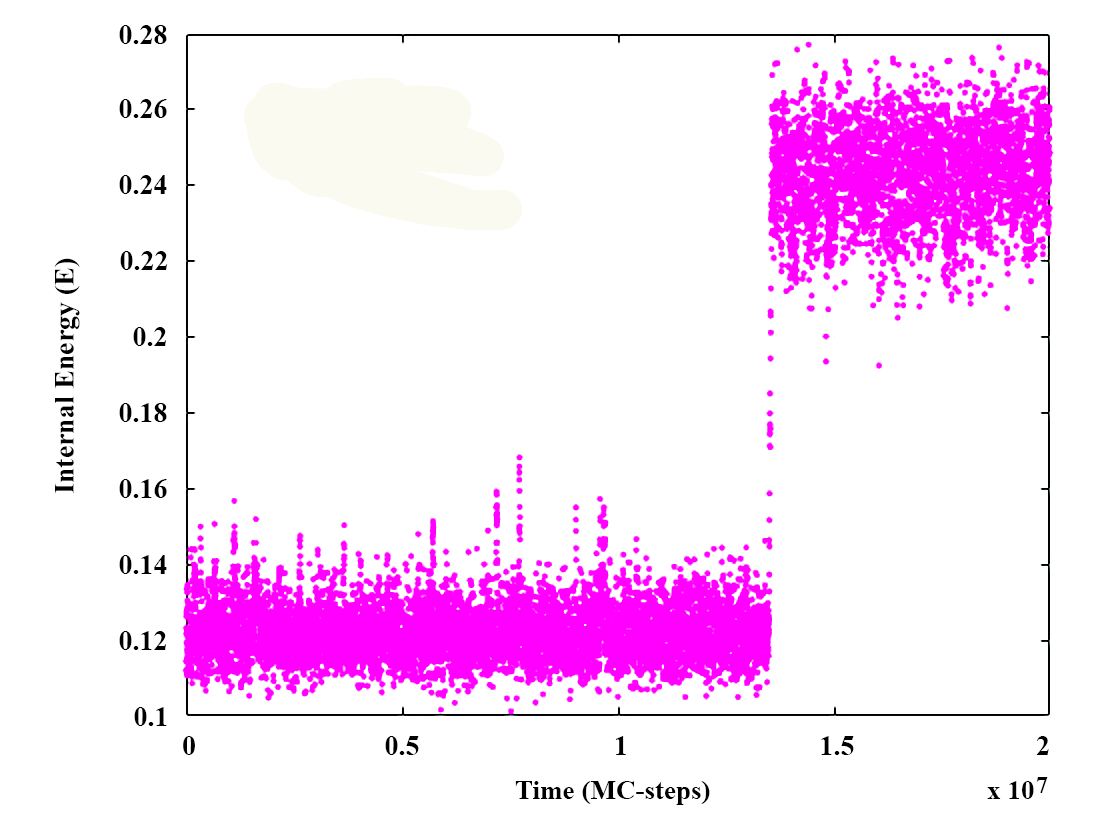}}
    \subfigure[]{\includegraphics[width=0.295\textwidth]{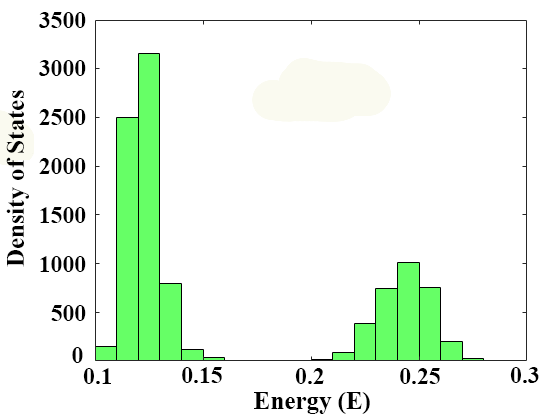}}
    \subfigure[]{\includegraphics[width=0.385\textwidth]{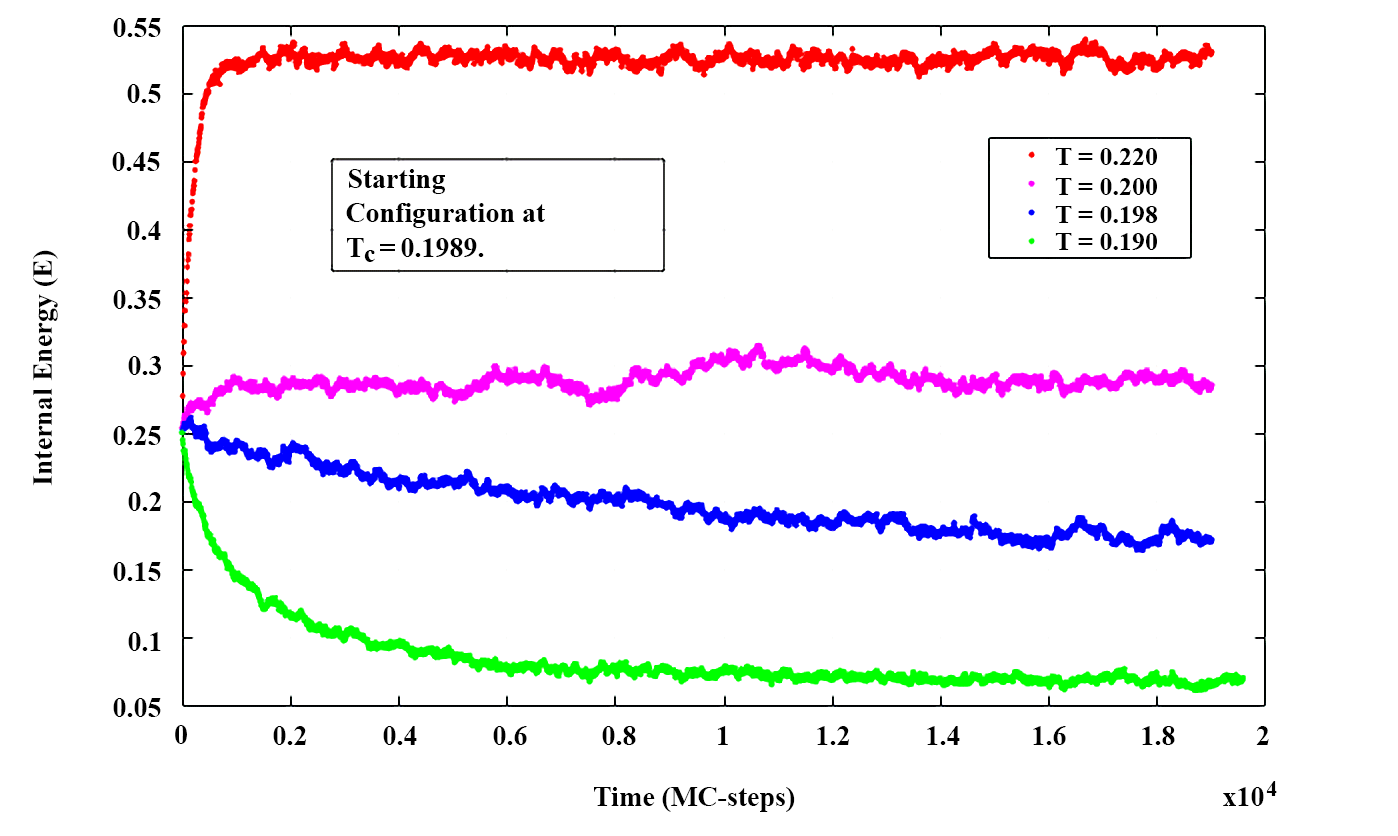}}
    \subfigure[]{\includegraphics[width=0.42\textwidth]{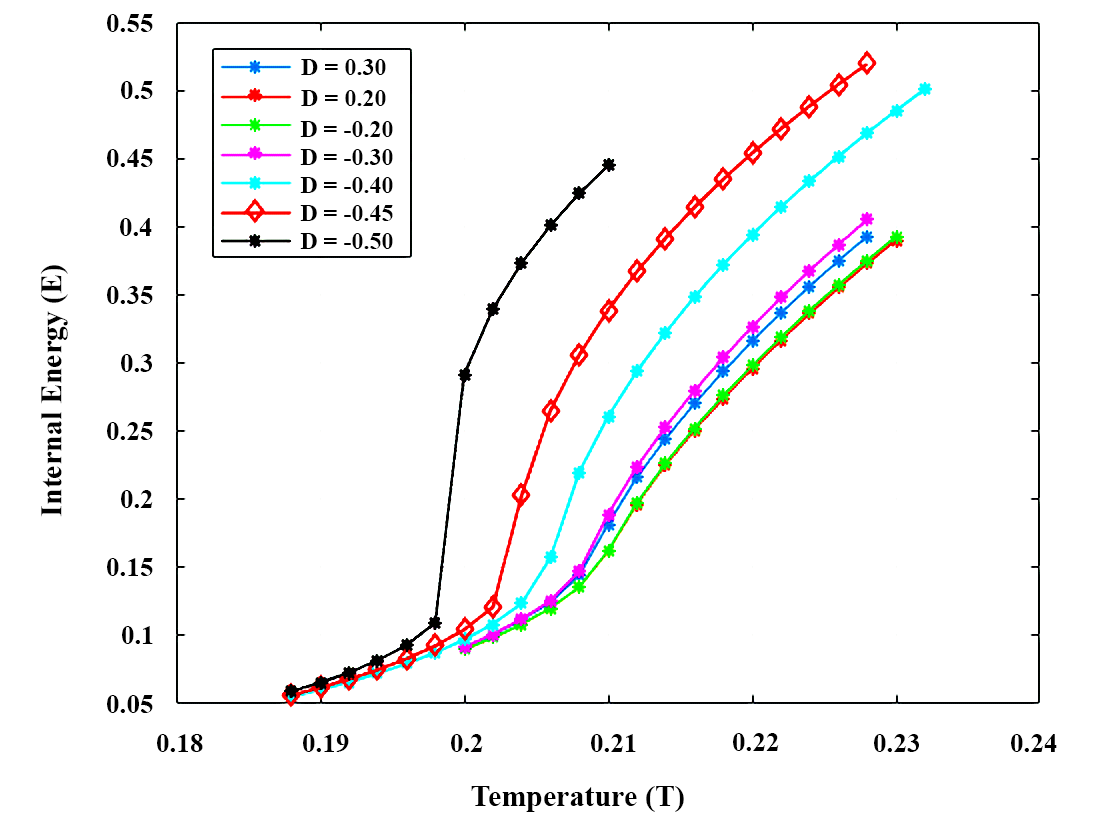}}
    \subfigure[]{\includegraphics[width=0.42\textwidth]{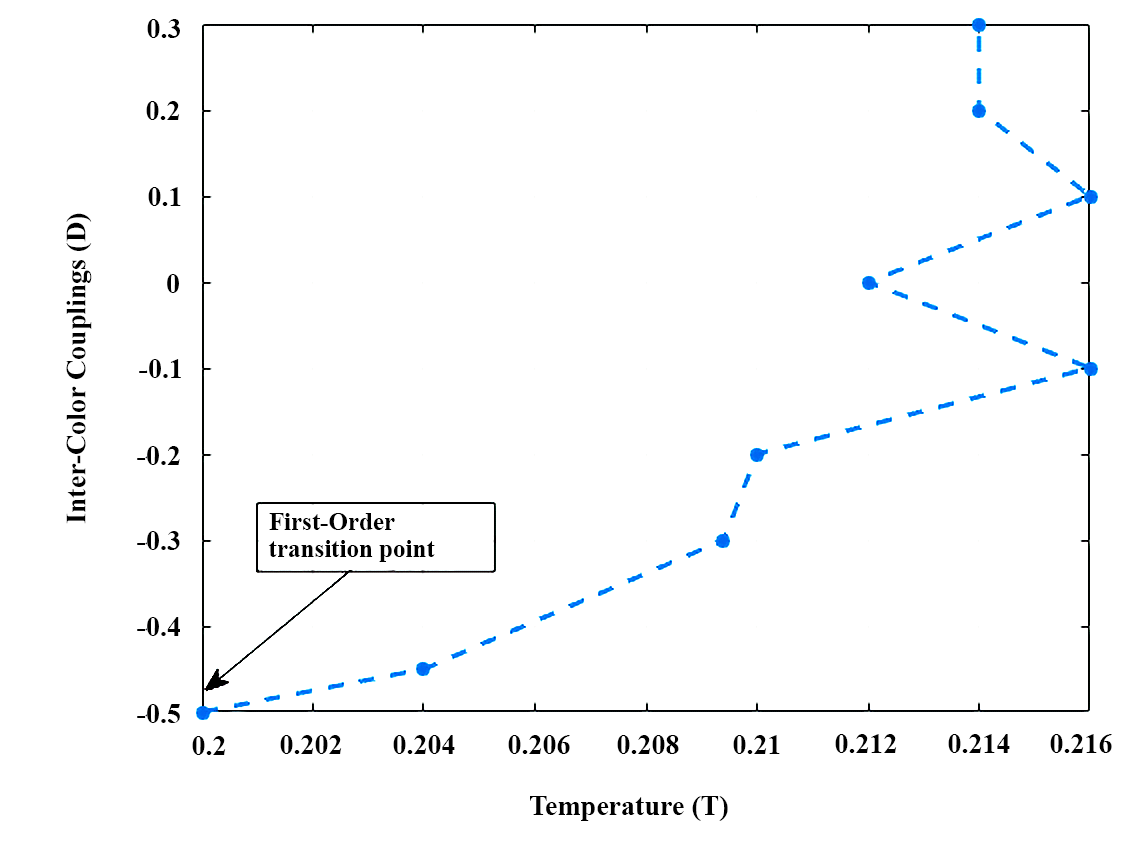}}
	\caption{Metastable behaviour in NS loops with $D = -0.5$ and $L = 26$: (a) the $E$ vs time (MC-steps) plot at $T = 0.1989$; 
	(b) density of states for the system above and below $T_{c}$; (c) $E$ vs time plot for the system at temperatures above and below 
	$T_{c} = 0.1989$; (d) $E-T$ plot for varying $D$ values; (e) $T$ vs $D$ plot (phase diagram) indicating the tricritical point.}
    \label{NSLoopsFirstOrder}
\end{figure}

In the previous section, we have seen that $\alpha$ changes with $D$. When the effect of inter-color coupling is no longer small, 
the nature of the transition in NS loops changes drastically. The $E$-$T$ plot in Fig.\ref{NSLoopsFirstOrder}(d) shows a 
reduction in the transition temperature for increasing negative values of $D$. Particularly for $D = -0.5$, the system jumps into the 
disordered state discontinuously. The critical temperature for this transition $T_{c} = 0.1989$ (Fig.\ref{NSLoopsFirstOrder}(a)). 
To study the metastability in NS loops for $D = -0.5$, we have repeated the procedure we did for the SY loops. The density 
of states for this system is found to have two peaks. The values of the internal energies were centred about $E{1} \approx 0.12$ and 
$E_{2} \approx 0.24$, respectively (Fig.\ref{NSLoopsFirstOrder}(a)). Using these values, the change in the latent heat for melting at 
$T_{c}$ is $\Delta E = E_{2} - E_{1} \approx 0.12$. Subsequently, the change in entropy is

\begin{equation}
\Delta S = \frac{\Delta E}{T_{c}} \approx 0.6(0332).
\end{equation}

Starting from a mixed state configuration, we evolve the system at temperatures both above and below $T_{c}$
(Fig.\ref{NSLoopsFirstOrder}(c)). Depending upon the temperature at which the simulations are performed, the system 
transits either into a high or a low-temperature phase. We have presented the $T$ vs $D$ phase diagram plot for this system 
in Fig.\ref{NSLoopsFirstOrder}(e). It shows the presence of a \textsl{tri-critical} point in the range $-0.5 < D < -0.45$. 

\section{Summary}

In this work, we studied the order-disorder transitions mediated by loop models in a regular three-dimensional lattice. 
We found that the nature of the phase transition is intimately related to the nature of these loops. The loop model with symmetric 
loops shows a first-order phase transition, whereas the loop model with non-symmetric loops shows a second-order phase transition. 
Interactions among different colors play an important role in NS loops. Finite-size scaling studies for the NS loops suggest that 
the correlation length exponent $\nu$ matches with the three-dimensional XY model. However, the specific heat exponent for any 
$D \neq 0$ disagrees with the values reported in the literature. Its value obtained from MC studies of the three-dimensional 
XY model \cite{Campostrini} is found to be $\alpha/\nu \approx 0.0191$. This is significantly different from what we observe in our 
NS-loop system. Therefore, we infer that the inter-color couplings significantly affect the nature of the phase transition.

The NS loop model was written down in a rather heuristic manner. It remains unclear to us as to how such a loop model can be derived
from first principles. Therefore, it is important to provide a proper foundation for this model. In principle, one can find a correspondence 
between the NS loops and the well-known $N$-vector spin model. The $N$-vector spins in the high-temperature limit can serve as the dual model 
for the NS loop model at low temperatures. Using the perturbation series expansion method \cite{Wortis,Oitmaa}, every low-temperature excitation 
in the NS loops will have a high-temperature counterpart in the spin model. To obtain the correspondence between these two models, the 
respective weight factors for these excitations can be calculated (to be communicated later separately).
\section{Acknowledgement}
The authors acknowledge the financial support from the Council of Scientific \& Industrial Research (CSIR), the Indo-U.S. Science \& Technology
Forum (IUSSTF), India. Also, the computational support from the Supercomputer Education \& Research Centre is acknowledged. S. Mukherjee and S.K. Ganguly
thank Prof. N. Chandra in the Bioinformatics lab for partial computational support. C. Dasgupta thanks the Department of Science \& Technology (DST),
Government of India. S.K. Ganguly would like to thank the entire ONGIL technical team, Mr. Amit Kumar Patra, Prof. Banibrata Mukhopadhyay, and Prof. Indrajit 
Mukherjee for their constant technical and moral support. Finally, we would like to express our gratitude to all the anonymous referees for their valuable 
suggestions.
\section{Appendix}
In this appendix, we will write down the update equations for the elementary excitations in SY loops. We will demonstrate two 
such cases, while the remaining can easily be read off from the figures in Fig.\ref{SYLoopExcitations}. 
Using the symmetrization relation in Eq.\eqref{SYMRelationB}, and the update equations for the NS loops (e.g. Eq.\eqref{NSLcase123A}), 
we will have the following update equations for the SY loops shown in Fig.\ref{SYLoopExcitations}(a) (left).

\begin{eqnarray}\label{SYcase3}
& \bar{\eta}_{11}(\mathbf{x}) \rightarrow \bar{\eta}_{11}(\mathbf{x}) + 2 & \bar{\eta}_{12}(\mathbf{x}) \rightarrow \bar{\eta}_{12}(\mathbf{x}) - 1 \\  \nonumber
& \bar{\eta}_{21}(\mathbf{x}) \rightarrow \bar{\eta}_{21}(\mathbf{x}) - 1 & \bar{\eta}_{23}(\mathbf{x}) \rightarrow \bar{\eta}_{23}(\mathbf{x}) + 1 \\  \nonumber
& \bar{\eta}_{32}(\mathbf{x}) \rightarrow \bar{\eta}_{32}(\mathbf{x}) + 1 & \bar{\eta}_{13}(\mathbf{x}) \rightarrow \bar{\eta}_{13}(\mathbf{x}) - 1 \\  \nonumber
& \bar{\eta}_{31}(\mathbf{x}) \rightarrow \bar{\eta}_{31}(\mathbf{x}) - 1 & 
	\bar{\eta}_{12}(\mathbf{x} + \hat{x}) \rightarrow \bar{\eta}_{12}(\mathbf{x} + \hat{x}) + 1   \\  \nonumber
& \bar{\eta}_{21}(\mathbf{x} + \hat{x}) \rightarrow \bar{\eta}_{21}(\mathbf{x} + \hat{x}) + 1 & 
	\bar{\eta}_{23}(\mathbf{x} + \hat{x}) \rightarrow \bar{\eta}_{23}(\mathbf{x} + \hat{x}) - 2   \\  \nonumber
& \bar{\eta}_{32}(\mathbf{x} + \hat{x}) \rightarrow \bar{\eta}_{32}(\mathbf{x} + \hat{x}) - 2 & 
	\bar{\eta}_{13}(\mathbf{x} + \hat{x}) \rightarrow \bar{\eta}_{13}(\mathbf{x} + \hat{x}) + 1   \\  \nonumber
& \bar{\eta}_{31}(\mathbf{x} + \hat{x}) \rightarrow \bar{\eta}_{31}(\mathbf{x} + \hat{x}) + 1 & 
	\bar{\eta}_{11}(\mathbf{x} + \hat{y}) \rightarrow \bar{\eta}_{11}(\mathbf{x} + \hat{y}) - 2   \\  \nonumber
& \bar{\eta}_{13}(\mathbf{x} + \hat{y}) \rightarrow \bar{\eta}_{13}(\mathbf{x} + \hat{y}) + 1 & 
	\bar{\eta}_{31}(\mathbf{x} + \hat{y}) \rightarrow \bar{\eta}_{31}(\mathbf{x} + \hat{y}) + 1   \\  \nonumber
& \bar{\eta}_{11}(\mathbf{x} + \hat{z}) \rightarrow \bar{\eta}_{11}(\mathbf{x} + \hat{z}) - 2 & 
	\bar{\eta}_{12}(\mathbf{x} + \hat{z}) \rightarrow \bar{\eta}_{12}(\mathbf{x} + \hat{z}) + 1    \\  \nonumber
& \bar{\eta}_{21}(\mathbf{x} + \hat{z}) \rightarrow \bar{\eta}_{21}(\mathbf{x} + \hat{z}) + 1 &
	\bar{\eta}_{23}(\mathbf{x} + 2\hat{x}) \rightarrow \bar{\eta}_{23}(\mathbf{x} + 2\hat{x}) + 1  \\  \nonumber
& \bar{\eta}_{32}(\mathbf{x} + 2\hat{x}) \rightarrow \bar{\eta}_{32}(\mathbf{x} + 2\hat{x}) + 1 &  
	\bar{\eta}_{12}(\mathbf{x} + \hat{x} + \hat{y}) \rightarrow \bar{\eta}_{12}(\mathbf{x} + \hat{x} + \hat{y}) - 1 \\  \nonumber
& \bar{\eta}_{21}(\mathbf{x} + \hat{x} + \hat{y}) \rightarrow \bar{\eta}_{21}(\mathbf{x} + \hat{x} + \hat{y}) - 1 & \quad
	\bar{\eta}_{11}(\mathbf{x} + \hat{y} + \hat{z}) \rightarrow \bar{\eta}_{11}(\mathbf{x} + \hat{y} + \hat{z}) + 2  \\  \nonumber
& \bar{\eta}_{12}(\mathbf{x} + \hat{x} + \hat{z}) \rightarrow \bar{\eta}_{12}(\mathbf{x} + \hat{x} + \hat{z}) - 1 & \quad
	\bar{\eta}_{21}(\mathbf{x} + \hat{x} + \hat{z}) \rightarrow \bar{\eta}_{21}(\mathbf{x} + \hat{x} + \hat{z}) - 1.
\end{eqnarray}

Note, that the symmetric nature of the loops is evident from the update equations. For example, the blue colored loop variable along the x-direction
$\eta_{12}(\mathbf{x})$ has a value $-1$ along the link $(\mathbf{x}, \mathbf{x} + \hat{x})$ (towards $\mathbf{x}$). The loop symmetric to this 
is the red colored loop, with value $-1$ along the link $(\mathbf{x}, \mathbf{x} + \hat{y})$. We can write similar update equations for the 
second excited state in Fig.\ref{SYLoopExcitations}(b) (left)

\begin{eqnarray}\label{SYcase4}
& \bar{\eta}_{11}(\mathbf{x})  \rightarrow  \bar{\eta}_{11}(\mathbf{x}) - 2 & 
	\bar{\eta}_{22}(\mathbf{x})  \rightarrow  \bar{\eta}_{22}(\mathbf{x}) - 2 \\  \nonumber
& \bar{\eta}_{12}(\mathbf{x})  \rightarrow  \bar{\eta}_{12}(\mathbf{x}) + 2 & 
	\bar{\eta}_{21}(\mathbf{x})  \rightarrow  \bar{\eta}_{21}(\mathbf{x}) + 2 \\  \nonumber
& \eta_{22}(\mathbf{x} + \hat{x})  \rightarrow  \eta_{22}(\mathbf{x} + \hat{x}) + 4 & 
	\eta_{12}(\mathbf{x} + \hat{x})  \rightarrow  \eta_{12}(\mathbf{x} + \hat{x}) - 2 \\  \nonumber
& \eta_{21}(\mathbf{x} + \hat{x})  \rightarrow  \eta_{21}(\mathbf{x} + \hat{x}) - 2 &  
	\eta_{11}(\mathbf{x} + \hat{y})  \rightarrow  \eta_{11}(\mathbf{x} + \hat{y}) + 4 \\  \nonumber
& \eta_{12}(\mathbf{x} + \hat{y})  \rightarrow  \eta_{12}(\mathbf{x} + \hat{y}) - 2 &  
	\eta_{21}(\mathbf{x} + \hat{y})  \rightarrow  \eta_{21}(\mathbf{x} + \hat{y}) - 2 \\  \nonumber
& \eta_{22}(\mathbf{x} + 2\hat{x}) \rightarrow  \eta_{22}(\mathbf{x} + 2\hat{x}) - 2 & 
	\eta_{11}(\mathbf{x} + 2\hat{y}) \rightarrow  \eta_{11}(\mathbf{x} + 2\hat{y}) - 2 \\  \nonumber
& \eta_{12}(\mathbf{x} + \hat{x} + \hat{y}) \rightarrow \eta_{12}(\mathbf{x} + \hat{x} + \hat{y}) + 2 & \quad
	\eta_{21}(\mathbf{x} + \hat{x} + \hat{y}) \rightarrow \eta_{21}(\mathbf{x} + \hat{x} + \hat{y}) + 2.
\end{eqnarray}

Following the above procedures, the update equations for the other excitations can be written down easily.  


\end{document}